\documentclass[12pt]{article}
\pdfoutput=1
\usepackage[totalwidth=480pt,totalheight=640pt]{geometry}
\usepackage[centertags]{amsmath}
\usepackage[square, comma, sort&compress,numbers]{natbib}
\usepackage{array,multirow}
\numberwithin{equation}{section}
\usepackage{amssymb}
\usepackage{graphicx}
\usepackage{color}
\usepackage{setspace}

\usepackage{epsfig}

\def\la{\lambda}

\def\Or[#1]{{\text{O}}\left({#1}\right)}
\def\dotl[#1,#2]{\left\langle #1, #2 \right\rangle}
\def\dotlb[#1,#2]{[ #1, #2 ]}
\def\dotp[#1,#2]{(#1) \cdot (#2)}
\def\aff[#1,#2]{\hat{#1}(#2)}
\def\n4sym{{\cal N}=4 SYM}
\def\>{\rangle}
\def\<{\langle}
\def\weight[#1,#2,#3]{\{(#1),#2,#3\}}
\def\ads[#1]{$\text{AdS}_{#1}$}

\newcommand{\ba}{\begin{eqnarray}}
\newcommand{\ea}{\end{eqnarray}}
\newcommand{\be}{\begin{eqnarray}}
\newcommand{\ee}{\end{eqnarray}}
\newcommand{\bq}{\begin{equation}}
\newcommand{\eq}{\end{equation}}

\newcommand{\CA}{{\cal A}}

\newcommand{\CC}{{\cal C}}
\newcommand{\CD}{{\cal D}}

\newcommand{\CF}{{\cal F}}

\newcommand{\CO}{{\cal O}}

\newcommand{\nn}{\nonumber}

\newcommand\oo\infty
\newcommand\s\sigma
\newcommand\de\delta
\newcommand\De\Delta

\newcommand\f\phi
\newcommand\g\gamma
\newcommand\x\times

\newcommand{\ra}{\rightarrow}

\newcommand{\sfrac}[2]{{\textstyle\frac{#1}{#2}}}
\newcommand\pd{\partial}

\newcommand\mpld{M_{\rm {Pl}, d+1}}
\newcommand\mplthr{M_{\rm {Pl}, 3}}

\newcommand\na{\nabla}

\renewcommand\la{\langle}
\renewcommand\ra{\rangle}
\newcommand\rmd{{\rm d}}

\newcommand\zUV{z_{\rm UV}}
\newcommand\HIR{H_{\rm IR}}
\newcommand\HUV{H_{\rm UV}}
\newcommand\vep{\varepsilon}
\newcommand\rUV{r_{\rm UV}}

\numberwithin{equation}{section}

\usepackage{hyperref}

\begin{document}

\begin{titlepage}

\begin{center}
\vspace{1cm}

{\Large \bf  An Effective Theory for Holographic RG Flows}

\vspace{0.8cm}

\small
\bf{Jared Kaplan,  Junpu Wang}
\normalsize

\vspace{.5cm}

{\it  Department of Physics and Astronomy, Johns Hopkins University, Baltimore, MD 21218} 

\end{center}

\vspace{1cm}

\begin{abstract}

We study the dilaton action induced by RG flows between holographic CFT fixed points.  For this purpose we introduce a general bulk effective theory for the goldstone boson of the broken spacetime symmetry, providing an AdS analog of the EFT of Inflation.  In two dimensions, we use the effective theory to compute the dilaton action, as well as the UV and IR conformal anomalies, without further assumptions.  In higher dimensions we take a `slow-flow' limit analogous to the assumption of slow-roll in Inflation, and in this context we obtain the dilaton action, focusing on terms proportional to the difference of the A-type anomalies.  We include Gauss-Bonnet terms in the gravitational action in order to verify that our method correctly differentiates between A-type and other anomalies.

\end{abstract}

\bigskip

\end{titlepage}

\tableofcontents

\section{Introduction }
\label{sec:Introduction}

Quantum Field Theories (QFT) can be defined via Renormalization Group (RG) flows between Conformal Field Theory (CFT) fixed points.  The QFT action can be written as the sum of a high-energy CFT action, $S_{CFT}$, plus a set of relevant operators $O_i$ with coefficients  $\phi^{(0)}_i$ defined at the high-energy scale $\Lambda$.  Major insight can be gained by promoting this unique scale $\Lambda \to \Lambda e^{-\tau(x)}$ to a spacetime dependent background dilaton field \cite{Komargodski:2011vj, Komargodski:2011xv, Luty:2012ww, Dymarsky:2013pqa}.   The effective $\tau$ action is constrained by symmetry, and certain terms can be determined by anomaly matching.  Alternatively, the full action may be directly computed in particular examples.  

In this note we will study holographic renormalization flows by erecting a general holographic effective field theory for $\pi$, the goldstone boson of the broken spacetime symmetry dual to dilatations.  Thus our approach may be viewed as intermediate between a general conformal symmetry analysis \cite{Imbimbo:1999bj, Schwimmer:2003eq, Schwimmer:2013jma, Cabo-Bizet:2013lha} and the study of particular holographic examples (see \cite{Skenderis:2002wp} for a review and many older references, and also the recent work \cite{Hoyos:2012xc, Hoyos:2013gma, Kol:2013msa}) for the study of the low-energy $\tau$ action.  We will be able to determine universal properties of this action by constructing a bulk $\pi$ action and solving the equations of motion for $\pi$.  A fairly complete analysis is possible for 2d QFTs, but in general dimensions more involved computations are required.  In a `slow-flow' limit where one can neglect mixing with AdS gravity, the leading low-energy $\tau$ action can still be determined in general $d$.  We assume that the bulk dynamics satisfy the null energy condition (NEC), which implies holographic  \cite{Myers:2010tj} $c$ or $a$ theorems  \cite{Zamolodchikov:1986gt, Cardy:1988cwa}, in two or higher dimensions, respectively. 

For notational and conceptual convenience, we display an explicit Weyl factor in the boundary metric, so that $g_{\mu \nu} \to e^{2 \zeta(x)} \hat g_{\mu \nu}$, where $\hat g$ has fixed determinant.  Then we can conveniently study Weyl transformations $\zeta \to \zeta - \sigma$.  The explicit presence of $\zeta$ also enables a computation of the UV and IR CFT trace anomalies via differentiation with respect to $\zeta$;  in contrast the $\tau$ action is only directly sensitive to the difference between the UV and IR anomalies.  In fact, $\zeta$ and $\tau$ are closely related, because simultaneous Weyl transformations and shifts of $\tau$ remain an (anomalous) symmetry of the QFT \cite{Osborn:1993cr, Maldacena:2002vr, Komargodski:2011xv} in the presence of the conformal symmetry breaking couplings.  The background fields $\zeta$ and $\tau$ are not quite identical, due to the presence of conformal anomalies.

Both the physics and our notation borrow from a closely related theory of broken spacetime translations, namely the effective field theory of inflation \cite{EFTInflation}.  In that case it is the deSitter time-translation symmetry which is broken, whereas in our case it is the holographic radial direction in AdS, but the two are related by analytic continuations.  The interpretation of boundary conditions distinguishes the two cases conceptually \cite{Maldacena:2002vr}.  In inflation we first compute the wavefunction of the bulk fields, and then we compute fixed time correlators by multiplying the wavefunction by powers of these fields and integrating over them. In contrast, in AdS/CFT we prescribe fixed boundary conditions for the fields in the UV region, including the metric, in order to compute a generating function for CFT correlators.  These fixed UV boundary conditions are crucial for our EFT of renormalization flows, because they permit a separation between the fluctuations of the boundary metric (i.~e.~the deviation from the flat metric) and those of the `matter' fields that produce conformal symmetry breaking.  In the presence of anomalies, this differentiates the dilaton $\tau$ from the trace of the metric, parameterized by the scalar field $\zeta$.

Although our treatment of the 2d holographic conformal anomaly will be general, in higher dimensions we make assumptions about the breaking of conformal symmetry -- we assume that the UV CFT is perturbed by only a single relevant operator, and that the slow renormalization flow is a small perturbation of the CFT, as we make precise in section \ref{sec:GeneralDimensions}.  In the future it would be interesting to provide a more complete demonstration of the universality of the conformal anomalies in a holographic context.  In the cases we study, the universality of the anomaly terms follows because the bulk $\pi$ action can be determined via a matching procedure sensitive only to the UV and IR regions of the bulk.  Roughly speaking, this can be viewed as a generalization of the Israel junction condition for a domain wall.

 This paper is organized as follows.  In section \ref{sec:Flows} we review the background dilaton formalism and its relationship with conformal anomalies, and then recast the discussion in a holographic context, giving some simple examples of the bulk effective theory for $\pi$.  Then in section \ref{sec:2dFlows} we study the 2d case in detail, obtaining the anomalies of the UV and IR CFTs and the $\tau$ action from holography.  In section \ref{sec:GeneralDimensions} we study the higher dimensional case by making more restrictive slow-flow and demixing assumptions, but including a Gauss-Bonnet term to show how we distinguish the A-type anomalies.  In section \ref{sec:Caveats} we consider two more complicated scenarios, involving higher derivative terms in the bulk action and multiple relevant operators perturbing the UV CFT, and show that in the limits we have conisidered, these effects do not alter our results.  In section \ref{sec:Discussion} we discuss the results.  In appendix \ref{eq:RigorousActionComputations} we provide more detailed and thorough calculations based on a solution matching method, and in appendix \ref{sec:AxialGauge} we show how the $\tau$ and $\zeta$ actions can be determined in axial gauge.  In section \ref{sec:Caveats} we discuss the inclusion of higher derivative operators in the bulk and the case of multiple bulk fields, which is dual to a simultaneous perturbation of the UV CFT by several relevant operators.  Throughout this paper, we will use the Euclidean signature. We will use the Greek letters $\mu, \nu, \dots$ to denote the bulk coordinates, while the lowercase Latin letters $i,j,\dots$ to denote the boundary coordinates.

\section{Spurion Fields and Holographic Flows }
\label{sec:Flows}

Even when symmetries are broken, we may nevertheless pretend otherwise.  This is the idea behind the spurion method, which promotes symmetry breaking coupling constants to spacetime-dependent fields.  The transformations of the fields restore the symmetry, which then constrains the coupling dependence of physical observables.  Let us review this method as it has been applied to the breaking of conformal symmetry \cite{Komargodski:2011vj, Komargodski:2011xv, Elvang:2012yc} by renormalization flows.

In the case of QFTs flowing between a UV and IR CFT, we can characterize the high-energy theory with an action $S_{CFT}$ perturbed by various operators $O_i$, yielding the full action
\be
\label{eq:QFTAction}
S = S_{CFT}[g] + \sum_{n}\int d^d x \sqrt{g} \phi_n^{(0)}(x) O_n(x)
\ee  
where $\phi_n^{(0)}$ are effective coupling constants.  The perturbation will generically break the conformal symmetry, but we can restore it if we let the $\phi_n^{(0)}$ transform.  Specifically, following \cite{Komargodski:2011xv} let us replace every mass scale appearing in the couplings via $M \to M e^{-\tau(x)}$; this includes both explicit scales and implicit scales used to define the couplings.  

Weyl transformations act on the background metric $g_{\mu \nu}$ and the dilaton $\tau(x)$ as
\be
\label{eq:WeylTransform}
g_{i j} \to e^{-2 \sigma(x)} g_{ ij } \ \ \ \mathrm{and}  \ \ \ \tau(x) \to \tau(x) + \sigma(x)
\ee
The action in equation (\ref{eq:QFTAction}) will be invariant under these combined transformations, up to trace anomalies.  In fact, anomalies differentiate the Weyl factor of $g_{ij}$ from $\tau$, and this subtlety will be important in the holographic setup to be discussed below.  
We can use this invariance to constrain the construction of physical observables, and properties of the induced $\tau$ action provide interesting information about the renormalization flow.  In cases where conformal symmetry is spontaneously broken, the $\tau$ field will correspond with a physical degree of freedom, the massless goldstone boson associated with the spontaneous breaking.

These ideas have been used to great effect \cite{Komargodski:2011vj, Komargodski:2011xv, Luty:2012ww, Dymarsky:2013pqa} in the study of renormalization flows, where anomalies, including the conformal anomaly coefficients, must match between the UV and IR theories.  Let us first consider the $d=2$ case. 
In a curved background, there is a conformal symmetry violating trace anomaly\footnote{Our convention is the same as that in Ref.~\cite{Elvang:2012yc}, in which the trace anomaly for a $d=2p$ dimensional CFT in a curved background can be written as 
\be
\la T^j_{\; j} \ra = \sum_{i} c_i I_i- a(-1)^{d/2} E_{d}+B' \na_j J^j\;, 
\ee
where $c_i$'s, the coefficients for Weyl invariants $I_i$, are the central charges of the CFT and $a$ the ``type A'' anomaly. We normalize the Euler density in $d=2p$ dimension as 
\be
E_{2p}=\frac{1}{2^p} R_{\mu_1 \nu_1}^{\quad\;\;\rho_1\sigma_1}
\dots R_{\mu_p \nu_p}^{\quad\;\;\rho_p\sigma_p}\epsilon_{\rho_1\sigma_1\dots \rho_p \sigma_p}\epsilon^{\mu_1 \nu_1\dots \mu_k \nu_k}\;.
\ee}
\be
\label{eq:2dTraceAnomaly}
T^i_{\;\;i} = -\frac{c}{24 \pi} R
\ee
with $c$ being the central charge of a two-dimensional CFT. 
In fact we can construct a c-function along the RG flow, which is defined for all energy scales and is equal to $c_{\rm UV}$ and $c_{\rm IR}$, respectively, at the UV and IR fixed point. To state this in a bit more detail, note that we can write \cite{Cappelli:1990yc}
\be\label{eq:TT correlator}
\la T^i_{\;\;i}( k)T^j_{\;\;j}(-k)\ra=\frac{1}{12\pi}\int_0^{\infty}\!\rmd \mu \,c(\mu)\,\frac{(k^2)^2}{k^2+\mu^2}\;.
\ee
The function $c(\mu)$ is the spectral density, which is non-negative for all energy scales. It can be used to construct the c-function and to prove Zamolodchikov's c-theorem \cite{Zamolodchikov:1986gt, Komargodski:2011xv}. In particular, in QFTs with both UV and IR fixed points, it takes the form of 
\be
c(\mu)=c_{\rm IR}\delta(\mu)+c_1(\mu, \Lambda)\;,
\ee
where $c_1$ has a support away from $\mu=0$ and depends on the scale of conformal symmetry breaking.
The central charges of UV and IR CFT are given in terms of this spectral density $c(\mu)$ via the following integral representations: 
\be
c_{\rm UV}=\int_0^{\infty}\!\rmd \mu\,c(\mu)\;,\quad c_{\rm IR}=\lim_{\epsilon\to 0}\int_0^{\epsilon}\!\rmd \mu\,c(\mu)\;.
\ee
In the limit $k\to 0$, only the delta function support contributes to the integral in equation (\ref{eq:TT correlator}), and so we find 
\be\label{TT IR}
\lim_{k\to 0}\,\la T^i_{\;\;i}( k)T^j_{\;\;j}(-k)\ra= \frac{c_{\rm IR}}{12\pi} k^2\;,
\ee
 while at large $k$ the entire integral of $\mu$ contributes and we obtain 
 \be\label{TT UV}
 \lim_{k\to \infty}\,\la T^i_{\;\;i}( k)T^j_{\;\;j}(-k)\ra= \frac{c_{\rm UV}}{12\pi} k^2\;.
 \ee

The central charges for the UV and IR CFT, in general, are not equal. This means that the low-energy effective action for $\tau$ will not be entirely invariant under Weyl transformations, as it will be needed to compensate for the discrepancy between the UV and IR anomalies.  

To be explicit, we consider the generating function of the boundary QFT defined  by 
\be\label{generating function bdy}
e^{W_{\rm QFT} [g_{(0)},\tau]} \equiv 
\int\! {\cal D}\Psi_{\rm CFT} \exp\Bigg(- S_{CFT}[g_{(0)}, \Psi] - \sum_{n}\int d^d x \sqrt{g_{(0)}} \phi_n^{(0)}(x, \mu e^{-\tau}) O_n(x)\Bigg),
\ee
where $g_{(0)}$ is the metric on the boundary and $\mu$ is an arbitrary reference scale. For simplicity, let us assume that the boundary metric takes the form $g_{(0)i j }=e^{2\zeta}\delta_{ij}$, which can always be achieved in 2d via a boundary diffeomorphism. The Weyl transformation \eqref{eq:WeylTransform} reduces to 
\be\label{reduce Weyl}
\zeta\to \zeta-\sigma\;,\quad \tau \to \tau+\sigma\;.
\ee
Since the $O_n$'s are relevant operators in the UV CFT, $W_{\rm QFT}$ will be independent of $\tau$ at sufficiently short distances.  Thus the generating function $W_{\rm QFT}$ must take the form 
\be
W_{\rm QFT}[\zeta, \tau] \to\frac{c_{\rm UV}}{24\pi} \int\! d^2 x (\pd \zeta)^2\;,
\ee
in the UV limit, where the momenta $k$ are much larger than the scales associated with the relevant operators,
so that $\la T^i_{\;\;i}(\vec k)T^j_{\;\;j}(-\vec k)\ra =\frac{\delta^2 W_{\rm QFT}}{\delta \zeta(\vec k)\delta \zeta (-\vec k)}$ agrees with equation \eqref{TT UV}.  
At low energies the generating function must include the Wess-Zumino term for the dilaton field $\tau$: 
\begin{align}
W[\zeta, \tau] &\to \frac{c_{\rm IR}}{24\pi} \int\! d^2 x (\pd \zeta)^2+S_{\rm WZ}+\dots \label{IR W}\\
S_{\rm WZ}& ={\cal A}  \int \rmd^2 x  \sqrt{g} \left( \tau R  + (\nabla \tau)^2 \right)\label{2d WZ term}\;,
\end{align}
where dots denotes terms in higher order in derivatives. The first term on the right hand side of \eqref{IR W} reproduces \eqref{TT IR}, while the coefficient $\CA$ in the Wess-Zumino action can be determined via the anomaly matching, as we will explained below.

Notice that individual pieces of $S_{\rm WZ}$ transform as
\be
\sqrt{g}  \tau R &\to& \sqrt{g} e^{-2\sigma}  \left( \tau R  + 2 \tau \nabla^2 \sigma + \sigma R \right)
\\
\sqrt{g} (\nabla\tau)^2  & \to &   \sqrt{g} e^{-2\sigma}\left(  (\nabla \tau)^2  - 2 \tau \nabla^2 \sigma \right)
\ee
where we have performed an integration by parts in the second case.  
Therefore, under the infinitesimal Weyl transformation of equation (\ref{eq:WeylTransform}), the variation of $W_{\rm QFT}$ yields the anomaly of the theory, which must be scale-independent. Working up to quadratic order in the fields in the action, the Weyl variation is
\begin{align}
\frac{c_{\rm UV}}{12\pi}\int\! \rmd^2 x\, \sigma\,\pd^2 \zeta=\frac{c_{\rm IR}}{12\pi}\int\! \rmd^2 x\, \sigma\,\pd^2 \zeta
-2 \CA\int\! \rmd^2 x\, \sigma\,\pd^2 \zeta
\end{align}
where we have written the background curvature $R$ in terms of $\zeta$.
The left hand side of the above equation is the variation of the generating function evaluated near the UV fixed point where the dilaton field $\tau$ is absent, while the right hand side is near the IR fixed point.  
We conclude that ${\cal A}=\frac{c_{\rm UV}-c_{\rm IR}}{24 \pi}$.

Even when we take a flat metric, so that the Ricci scalar $R=0$,  $S_{\rm WZ}$ does not vanish, as we are left with the $(\partial \tau)^2$ term:
\be \label{eq:2dGoldstoneWZAction}
S_{\rm WZ} = - \frac{c_{\rm UV} - c_{\rm IR}}{24 \pi}  \int \rmd^2 x    (\partial \tau)^2 
\ee
The difference $c_{\rm UV} - c_{\rm IR}$ must be made up the Wess-Zumino term of equation (\ref{eq:2dGoldstoneWZAction}). 
 Equation \eqref{TT IR} and \eqref{TT UV} will be useful in comparison with the holographic computations.  

Conformal symmetry breaking will be geometrized in CFTs with holographic descriptions.  The conformal symmetries become the spacetime isometries of AdS spacetime, and so the breaking of conformal symmetry corresponds with the breaking of the AdS isometries by non-trivial bulk field configurations.  In the case of spontaneous breaking, where the coupling $\phi^{(0)}$ vanishes as we approach  the UV fixed point, the $\tau$ field will be related to the goldstone mode $\pi$ of the broken spacetime symmetry.  Even in the case of explicit breaking where $\phi^{(0)}$ remains nonzero at very high energy, $\tau$ plays the role of the spurion that non-linearly realizes the conformal symmetry.   Physically, if the $O_n$ are relevant operators in the UV CFT, at a sufficiently high energy scale the deformation of the CFT action (second term in \eqref{eq:QFTAction}) will always be subdominant. 

While $\tau(x)$ is manifested holographically as $\pi(r, x)$, the Weyl transformation of the $d$-dimensional background metric in equation (\ref{eq:WeylTransform}) can be realized holographically as well.  The fields $\zeta$ and $\tau$ are not quite equivalent, since the symmetry of equation (\ref{eq:WeylTransform}) is broken by the conformal anomaly.  Thus the low-energy $\zeta$ and $\tau$ actions differ, because $\zeta$ directly sources $T_i^i$ in the UV or IR CFT, while $\tau$ compensates for the difference in anomalies between these two theories.  We will see that the holographic actions for the trace of the bulk metric and for $\pi$ are also nearly identical, but differ by a crucial boundary term.

\begin{figure}[t!]
\begin{center}
\includegraphics[width=0.75\textwidth]{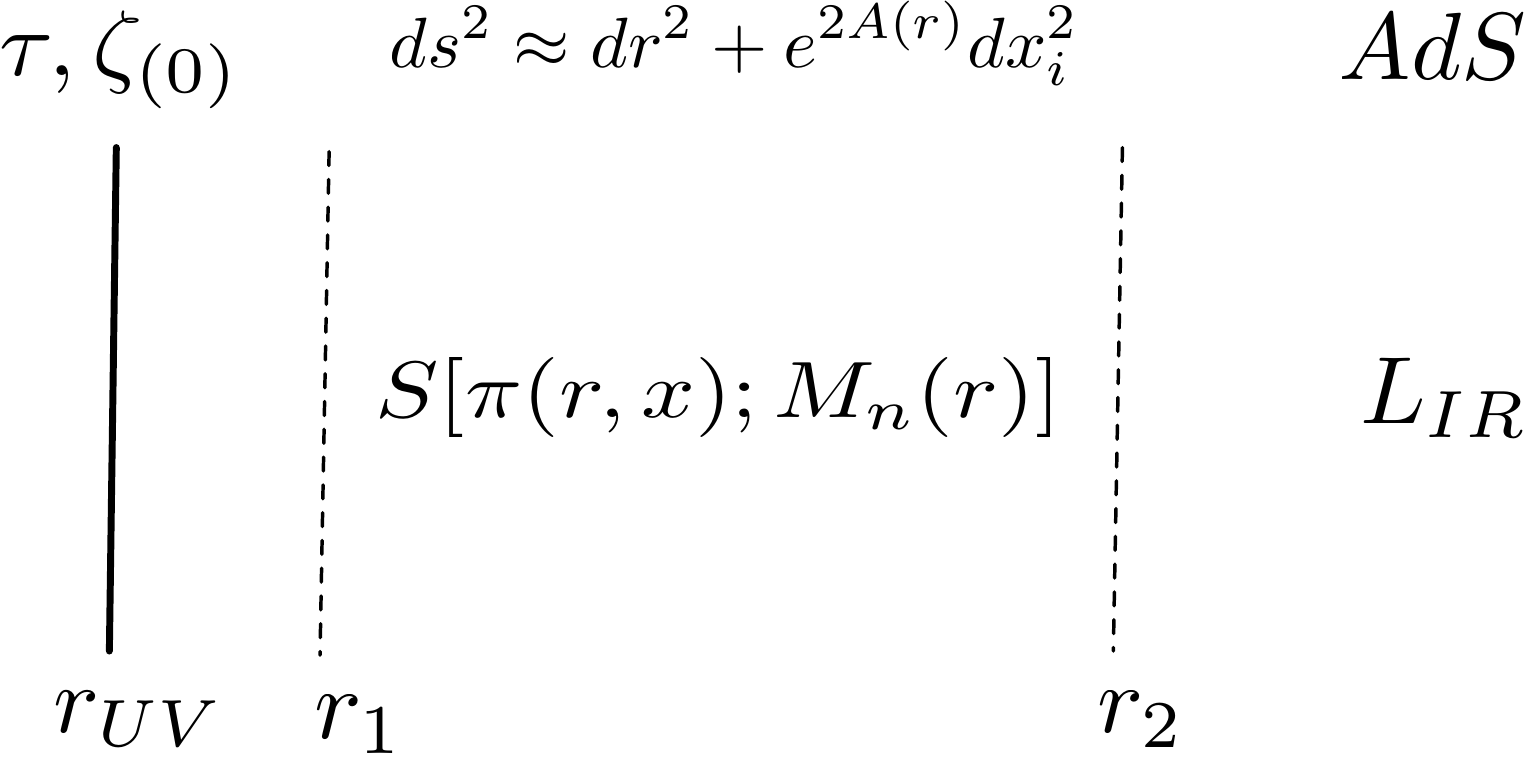}
\caption{ This figure indicates the holographic setup.  The background dilaton field $\tau$ in the CFT arises as the limit of the bulk goldstone field $\pi$ as it approaches the UV regulator surface at $r_{UV}$.  The bulk action for $\pi$ is only non-vanishing in the presence of the diffeomorphism breaking background parameterized by the $M_n(r)$.  The holographic RG flow ceases as we approach the deep IR, where we have a space with AdS length $L_{IR}$.   }
 \label{fig:SetupHolographicEFT} 
\end{center}
\end{figure}

We wrote the action  (\ref{eq:QFTAction}) in a form already suggestive of AdS/CFT.  In the illustrative case where the CFT lives in flat Euclidean space and where we neglect gravitational fluctuations for the moment, the bulk metric can be written as
\be
\rmd s^2 =  \rmd r^2 + e^{2 A(r)} \rmd x^i \rmd x_i
\ee
The couplings or sources $\phi^{(0)}_i$ are promoted to bulk fields $\phi_i(x, r)$ with boundary values $\phi^{(0)}_i$ at a UV regulator surface $r_{\rm UV}$, which we can later take to $+\infty$.  
In order for the bulk description to holographically describe a QFT flowing from a UV CFT to an IR CFT, we require
\begin{align}\label{IR UV Behavior}
\lim_{r\to r_{\rm UV}} A(r)=\frac{r}{L_{\rm UV}}\;,\quad \lim_{r\to -\infty} A(r)=\frac{r}{L_{\rm IR}}\;.
\end{align}
where $L_{\rm UV}$ and $L_{\rm IR}$ set the AdS scale, and therefore the central charge, of the UV and IR theories. Similarly, the scalar bulk fields $\phi_i$ asymptotes to fixed values in the UV region ($r\to r_{\rm UV}$) and the IR region ($r\to -\infty$).
  However, between these two extremes the function $A(r)$ and the fields $\phi_i$ can take any form consistent with their equations of motion, which follow from some bulk action $S_{\rm bulk}(g_{\mu \nu}, \phi_i)$.  In order to construct examples fulfilling these conditions, one must engineer the form of the bulk action (the potential for the $\phi_i$, for example). 

The  dilaton field $\tau$ can be introduced into the holographic description by promoting it to a bulk field $\pi(x, r)$ with boundary value $\pi(x, r_{\rUV}) = -L_{\rm UV}\tau(x)$.   The general setup is indicated in figure \ref{fig:SetupHolographicEFT}.  Note that according to equation (\ref{eq:WeylTransform}), $\tau(x)$ has the same conformal transformation properties as the AdS coordinate $r$.  This follows because dilatations in the CFT correspond to the AdS isometry $L_{\rm AdS}\partial_r + x^i \partial_{i}$.  So to implement adiabatic \cite{EFTInflation} fluctuations of the ``matter fields'' in the bulk, we write
\be 
\phi(x, r)\equiv \phi_{\rm bg}(r)+\delta \phi(x,r) =\phi_{\rm bg}(x, r + \pi(x, r))\;, 
\ee
then $\pi(x, r_{UV})$ becomes the desired collective deformation of the `couplings' $\phi^{(0)}_i(x) = \phi_i\big(x, \rUV+\pi(x, \rUV)\big)$, where we have interpreted the scale dependence of the couplings $\phi^{(0)}_i$ as arising from an $r_{UV}$-dependence of the bulk fields $\phi(x, r)$.  The fields that get $r$-dependent VEVs could even be composite operators in the bulk.  The fluctuations of the metric are parametrized by 
\be
\rmd s^2 =\rmd s_{\rm bg}^2+\delta g_{\mu\nu}\rmd X^\mu \rmd X^\nu\;.
\ee
 
 Physically, the non-trivial $r$-dependence of the bulk fields $\phi(x, r)$ encodes a breaking of the AdS isometries.  The  $\pi(x,r)$ field is a goldstone mode, so this breaking leads to a non-vanishing bulk action for $\pi(x,r)$.   A very similar theory for a spacetime goldstone mode was developed in the guise of the effective field theory of inflation \cite{EFTInflation}, based on the idea that slow-roll inflation involves a soft breaking of time translation symmetry.  The time coordinate in deSitter space can be viewed as the analytic continuation of the $r$ coordinate in AdS.  

The power of our method resides in the fact that it can be applied to completely general holographic RG flows.  Before we explain how to parameterize the general case, let us consider a simple example involving a single scalar field in the bulk with a canonical action
\be
S[\phi] = \int \!\rmd r \rmd^d x e^{d A(r)} \left( \frac{1}{2}  \nabla_\mu \phi \nabla^\mu \phi - V(\phi) \right)\;,
\ee
where the potential $V$ has at least two local minima, located at $\phi_{\rm UV}$ and $\phi_{\rm IR}$.
Subject to the boundary conditions
\be\label{BC for phi}
\lim_{r\to r_{\rm UV}} \phi_{\rm bg}(r)=\phi_{\rm UV}\equiv \phi^{(0)}\;,\quad \lim_{r\to -\infty} \phi_{\rm bg}(r)=\phi_{\rm IR}\;,
\ee
the scalar field will have some $x$-independent bulk profile $\phi_{\rm bg}(r)$.  The backreaction of the scalar field on the metric will also determine the function $A(r)$.  Note that the boundary conditions \eqref{BC for phi} insure that the bulk geometry asymptotes to pure AdS, since in the UV and the IR region the potential contributes as an effective cosmological constant, with the value $V(\phi_{\rm UV})$ and $V(\phi_{\rm IR})$, respectively,   which will determine the asymptotic AdS scale $L_{\rm UV}$ and $L_{\rm IR}$.  One can distinguish between spontaneous and explicit conformal symmetry breaking \cite{Klebanov:1999tb} by studying whether $\phi \to \phi_{UV}$, and at what rate.

 \begin{figure}[t!]
\begin{center}
\includegraphics[width=0.75\textwidth]{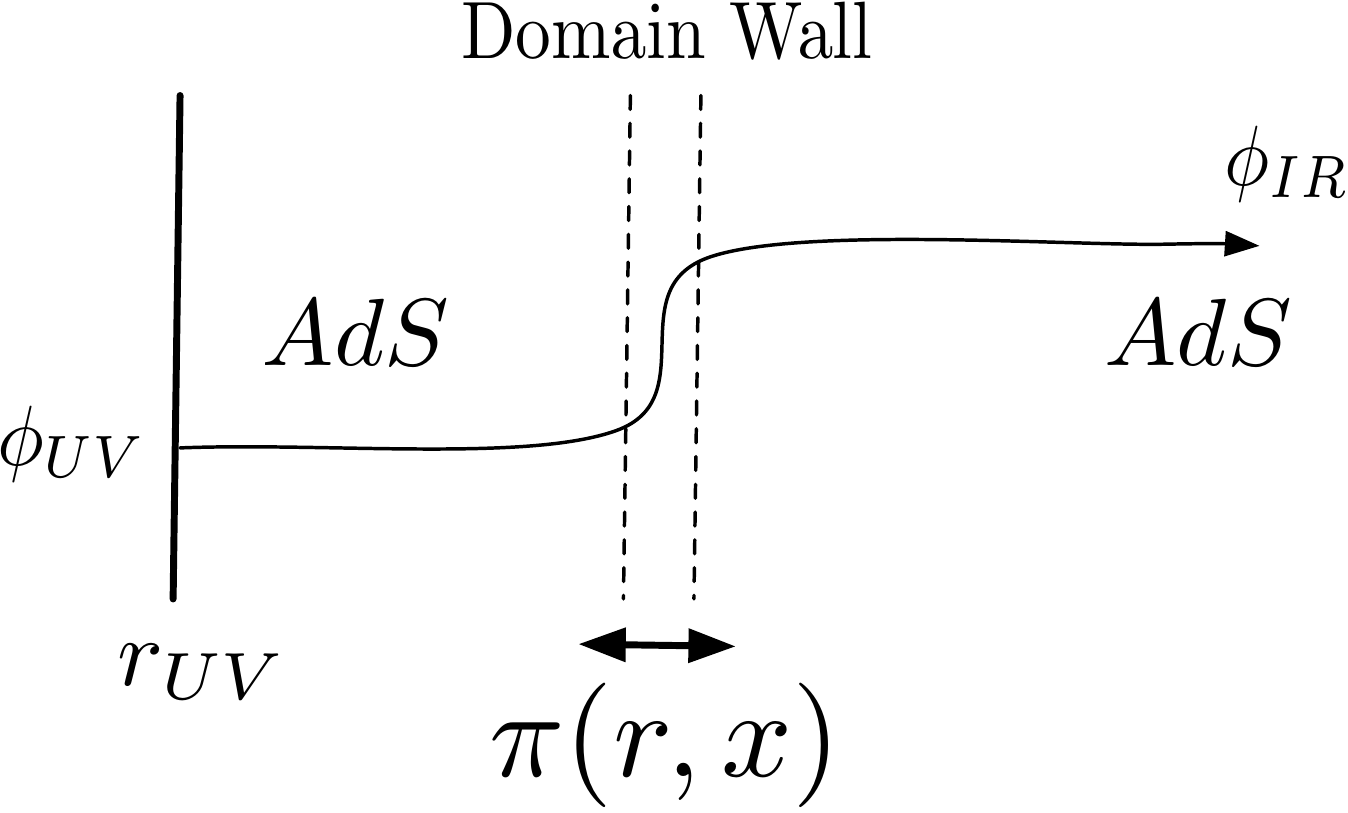}
\caption{ This figure indicates an extreme limit, where the RG flow of the CFT occurs in a narrow range of scales, corresponding to a thin domain wall in AdS.  In the long-wavelength limit, the $\pi$ field is a goldstone mode for fluctuations of the symmetry breaking domain wall.  Note that the Israel junction condition relates the tension of the domain wall with the change in the UV and IR cosmological constants.}
 \label{fig:DomainWall} 
\end{center}
\end{figure}
 
Now we can obtain an action for $\pi(x,r)$ via
\be
\label{eq:SimpleScalarExample}
S[\pi] &=& \int\! \rmd r \rmd^d x e^{d A(r)} \left( \frac{1}{2} \nabla_\mu \phi_{\rm bg}(r + \pi) \nabla^\mu \phi_{\rm bg}(r + \pi) - V(\phi_{\rm bg}(r + \pi)) \right) \nn
\\ & \approx & \int \!\rmd r \rmd^d x e^{d A(r)} \left( \frac{1}{2} \dot{\phi}_{\rm bg}^2  (\pd\pi)^2 - V(\phi_{\rm bg}(r + \pi)) \right)
\ee
A thin domain wall in AdS provides an extreme example of these ideas, and in such a case $\pi(x, r)$ would directly encode fluctuations of the wall.  This situation is pictured in figure \ref{fig:DomainWall}.  To illustrate it in more detail, let us specify
\be\label{thin wall phi}
\phi_{\rm bg}(r)=\phi_* \tanh\left(\frac{r-r_*}{w}\right)\;, \quad w\to 0
\ee
where $r_*$ is the position of the wall and $w$ is the width. Plugging this profile into \eqref{eq:SimpleScalarExample}, we find the kinetic term of $\pi$ takes the form
\begin{align}\label{thin wall kin}
S_{Kin}[\pi]&=\frac{\phi_*^2}{2w^2}\int \!\rmd r \rmd^d x e^{d A(r)} {\rm sech}^4\left(\frac{r-r_*}{w}\right) (\pd\pi)^2 \nn\\
&\simeq \frac{\phi_*^2}{2w }\int \!\rmd^d x e^{d A(r_*)} (\pd\pi)^2\;,
\end{align}
where in the second equality we have used the fact that in the limit $w\to 0$, $\dot{\phi}_{\rm bg}\propto {\rm sech}^2(\frac{r-r_*}{w})$ has a localized support in the vicinity of $r=r_*$.  The prefactor of the kinetic term can be interpreted as the domain wall tension.

An intuitive example in the opposite limit would be an extremely thick `domain wall', where $\dot{\phi}_{\rm bg}\ll \phi_{\rm bg} \dot{A}$. This is the AdS analogue \cite{Kol:2013msa} of slow roll inflation \cite{Maldacena:2002vr, EFTInflation}. We will make frequent use of this setup in our future discussions, especially when we study QFTs in dimension $d > 2$, since it will simplify the computations considerably. The procedures we have outlined above can be easily generalized to include higher order derivative terms in the $\phi$ action, or multiple bulk fields, as we discuss in section \ref{sec:Caveats}.

 \begin{figure}[t!]
\begin{center}
\includegraphics[width=0.75\textwidth]{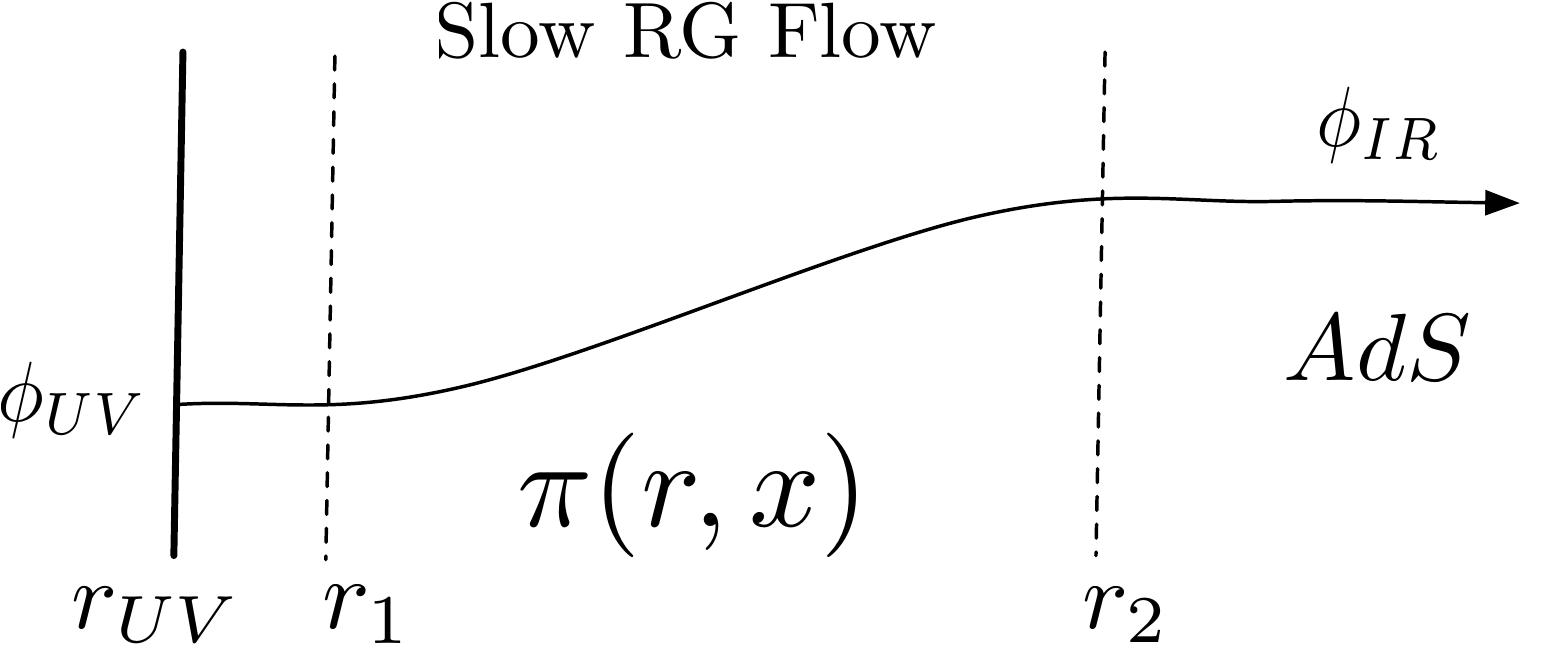}
\caption{ This figure suggests a limit where the RG flow in the QFT occurs `slowly' over a large range of scales.  This is an analog of slow roll inflation. }
 \label{fig:SlowFlow} 
\end{center}
\end{figure}

In the usual top-down approach to the study of holographic RG flows \cite{Skenderis:2002wp, Girardello:1998pd, Freedman:1999gp, Goldberger:1999uk, Hoyos:2012xc, Hoyos:2013gma, Kol:2013msa, Chacko:2013dra}, one needs to specify the bulk action and then solve the background equations of motions for the matter fields and the geometry.  The obvious drawbacks of such an approach are that we lack a general rule for what types of potentials and interactions to use in order to generate an RG flow for the boundary theory, and that i holographic RG models, the process of finding background profiles can be computationally cumbersome.  Our EFT approach, however, adopts a bottom-up viewpoint --- instead of exploring the space of bulk theories, we ask the question ``what is the most general bulk action describing the universal degrees of freedom in holographic RG flow.'' Like any EFT, our bulk action will contain undetermined (free) parameters, corresponding to the space of possible boundary QFTs.  As we will see in the following sections, our bulk action can reproduce  universal properties of RG flows from a UV CFT to an IR CFT.

Now let us construct a general bulk action for $\pi$.  As the goldstone boson of a broken spacetime symmetry, $\pi(x,r)$ necessarily mixes with the spacetime metric in the presence of dynamical gravity.  This means that in the bulk, we can obtain the action for $\pi$ via the `Stuckelberg trick' \cite{EFTInflation}.   
The idea is explained as follows: the bulk $r$-translation symmetry is spontaneously broken, due to the $r$ dependence in $\phi_{\rm bg}$ and in the background metric. We introduce $\pi$ in order to restore the bulk gauge redundancy under diffeorphisms; practically this means that we should write down terms in the bulk action that are invariant under the spatial diffeomorphism $x^i\to x^i+\xi^i(x,r)$, as well as the ``diagonal'' $r$-diffeomorphism $r \to r + \sigma(x, r),\; \pi\to \pi -\sigma (x,r)$.

At leading order in the derivative expansion, the bulk action is a sum of terms from the gravity sector, the matter sector, and from a counter-term action, although in general, gravity and matter cannot be separated.  We can write the action on the asymptotically AdS manifold ${\cal M}$ as 
\begin{align} 
S&=S_{\rm grav}+S_{\rm m}+S_{\rm ct}\;,\label{Bulk Action}\\
S_{\rm m}&=\sum_{n=0}^{\infty} \frac{1}{n!}\int\! \rmd r\,\rmd^{d} x \sqrt{g} M_{n}(r+\pi) Q^n \ + \ \text{Higher Derivatives} 
\label{Bulk Matter}
\end{align}
where we define
\be
Q \equiv \frac{\pd\big(r+\pi({\vec x},r)\big)}{\pd x^a}\frac{\pd\big(r+\pi({\vec x},r)\big)}{\pd x^b}g^{ab}({\vec x},r)\;.
\label{eq:QDef}
\ee
Note that the simple scalar example from equation (\ref{eq:SimpleScalarExample}) corresponds to the choice of parameters $M_2=M_3=\dots =0$.  By `higher derivatives' we indicate terms involving e.g. derivatives of the extrinsic curvature of constant $r$ slices, which we discuss in section \ref{sec:HigherDerivatives}.

We use Einstein gravity as $S_{\rm grav}$ in the case of 2d boundary QFTs,  but in $d\ge 4$ we study a gravity action with both an Einstein-Hilbert and a Gauss-Bonnet term, in order to distinguish the type A anomaly $a$ and the central charge $c$ of the asymptotic CFTs.  We include a UV boundary or regulator surface for the asymptotically AdS spacetime ${\cal M}$.   Therefore, to make the variational principle well defined, we include the Gibbons-Hawking-York boundary term in the gravitation action.  We also include counter terms on $\pd {\cal M}$  to cancel divergences \cite{Balasubramanian:1999re}. 

 Let us be explicit about our conventions concerning the bulk/boundary correspondence.  We have already defined the generating functional for the boundary QFT in equation \eqref{generating function bdy}. Meanwhile we also define the partition functional for the bulk theory to be
\be
Z_{\rm bulk}\equiv \int_{\phi_n^{(0)},\,g_{(0)}}{\cal D}\phi_{n}\CD g e^{- S_{\rm bulk}[g, \phi_n]}\simeq 
e^{-S_{\rm bulk}^{\rm on shell}[\phi_{\rm cl}, g_{\rm }]}\;,
\ee
In the second equality, we take the semi-classical limit by approximating the path integral with the classical bulk action evaluated on the classical solutions, subject to the boundary data $\phi_{(0)}$ and $g_{(0)}$.  The statement about correspondence is 
\be\label{duality}
W_{\rm QFT}=-S_{\rm bulk}^{\rm onshell}\;.
\ee 
so that we have $Z_{\rm QFT}=Z_{\rm bulk}$.

%%%%%%%%%%%%%%%%%%%%%%%%%%%%%%%
%%%%%%%%%%%%%%%%%%%%%%%%%%%%%%%%
\section{2d Conformal Anomaly and Holographic RG Flows}
\label{sec:2dFlows}

In this section we will give a general discussion of the 2d conformal anomaly, as it is obtained from holographic renormalization flows between CFT fixed points.  After setting up the problem in section \ref{sec:2dSetup}, we construct the bulk $\pi$ action in section \ref{sec:PiAction2d} and use it to compute the low-energy effective action for the dilaton $\tau$ at the quadratic level in derivatives and in $\tau$.  Then in section \ref{sec UV and IR Conformal Anomalies} we compute the action for $\zeta_{(0)}$, a Weyl factor for the boundary metric.  Differentiating the action with respect to $\zeta_{(0)}$ produces the $T_i^i$ correlators of equations (\ref{TT IR}) and (\ref{TT UV}) in the limit of large and small momenta, respectively.  Thus we obtain the UV and IR conformal anomalies, and the low energy action for the dilaton field $\tau$, which compensate the discrepancy between UV and IR conformal anomalies.  The bulk effective actions for $\pi$ and $\zeta$ are nearly identical, differing only by a total derivative in the bulk responsible for the conformal anomaly of the boundary QFT.

\subsection{The Setup}
\label{sec:2dSetup}

In this section, we take the gravity sector to have an action\footnote{The Planck mass in bulk dimension $d+1$ is related to the Newton constant by 
$\mpld^{d-1}=\left(8\pi G_N^{(d+1)}\right)^{-1}$.} 
\be
S_{\rm grav} =-\frac{\mplthr}{2} \int_{\cal M} R
+\mplthr\int _{\pd {\cal M}} K \label{2d Bulk Gravity}
\ee 
where 
$R$ is the $3d$ Ricci scalar, and $K$ is the trace of the extrinsic curvature of the boundary.  Einstein's equations for the background relate the coefficients $M_0$ and $M_1$ in the matter action in equation (\ref{Bulk Matter}), giving 
\begin{align}\label{M0M1}
M_0&=-\mplthr\left(\dot{H}+H^2\right)\nonumber\\
M_1&=-\frac{\mplthr}{2}\dot{H}\;.
\end{align}
These are directly analogous to the Friedman equations for the Hubble constant during inflation.
Here and henceforth we borrow the notation from cosmology by defining 
\be
a(r)=e^{A(r)}\;,\quad H(r)=\dot{A}(r)\;,\quad \dot{H}(r)=\ddot{A}(r)\;,
\ee
with dots denoting derivatives with respect to $r$.  We will also define
\be\label{eq:EpsilonCs}
\varepsilon & \equiv & -\frac{\dot{H}}{H^2} 
\nn \\
c_s^{-2} &\equiv&1-\frac{4 M_2(r)}{\dot{H}(r) \mplthr}\;.
\ee
We will see that $c_s$ is the relative normalization between gradient terms in the $r$ and $x^i$, so it would be the `speed of sound' in an analogue inflationary model.
There exists a `demixed' or `slow-flow' parametric limit of large $M_{\rm pl}$ with fixed matter energy density and slow $r$-variation where we can ignore the mixing of $\pi$ with gravity.  This is the limit where a goldstone equivalence theorem applies.  We will define this limit more precisely in subsequent sections, where it will be of use in studying higher dimensional examples.

In general we need to include gravitational effects, and this will not be prohibitively difficult for the case of 2d QFTs.  It is convenient to use the ADM variables to parametrize the Euclidean signature metric
\be
\rmd s^2=h_{ij}(N^i \rmd r+ \rmd x^i)(N^j \rmd r+ \rmd x^j)+N^2 \rmd r^2
\ee
where $h_{ij}$ is the induced metric on the constant $r$ slices and $h^{ij}$ is the inverse of the induced metric $h_{ij}$.  The inverse metric is
\be
g^{rr}=\frac{1}{N^2}\;, \quad g^{r i}=g^{i r}=-\frac{N^i}{N^2}\;, \quad g^{i j}=h^{i j}+\frac{N^i N^j}{N^2}.
\ee
In these variables we find
\be
Q=\frac{1}{N^2}\left(1+\pd_r \pi-N^i \pd_i \pi \right)^2+h^{i j}\pd_i \pi \pd_j \pi -1\;,
\ee
as defined in equation (\ref{eq:QDef}), and the gravitation action in \eqref{2d Bulk Gravity} becomes
\be
S_{\rm grav}=-\frac{\mplthr}{2}\int \! \rmd r\,\rmd^{d} x N\sqrt{h}\left(R^{(d)}+K^2-K^i_{\;\;j} K^j_{\;\;i}\right)\;,
\ee
where the extrinsic curvature terms are
\begin{align}
K_{ij}&=\frac{1}{2N}\left(\na_i N_j + \na_j N_i - \pd_r h_{ij} \right)\;, \nonumber\\
K&\equiv h^{ij}K_{ij}\;.
\end{align}
and the indices are raised and lowered with the induced metric $h_{ij}$. The lapse and shift function $N$ and $N^i$ are non-dynamical --- they can be algebraically determined in terms of $h_{ij}$. Moreover, the counter-term action in $d=2$ is given by \cite{Balasubramanian:1999re}
\be\label{2d counter}
S_{\rm ct}^{2d}=\frac{\mplthr}{ L_{\rm UV}}\int_{r=r_{\rm UV}}\! \rmd^2 x \sqrt{h} \;.
\ee

%%%%%%%%%%%%%%%%%%%%%%%%%%%%%
\subsection{Computing the $\pi$ Action}
\label{sec:PiAction2d}

Let us first study the simplest case, where the boundary metric is  flat  $h^{(0)}_{ij} = a(\rUV)^2\delta_{ij}$, and take the matter action in the bulk to be \eqref{Bulk Matter} while neglecting higher derivative terms. Since there are no tensor perturbations (i.e.~graviton degrees of freedom) in $3d$ gravity, we can gauge fix the bulk metric $h_{ij}$ so that it is flat everywhere:
\be
h_{ij}=a(r)^2 \delta_{ij}\;,\quad \phi(x, r)=\phi_{\rm bg}\big(r+\pi(x, r )\big)\;.
\ee
 This choice is consistent with our boundary condition for $h^{(0)}_{ij}$.  Since the only dynamical field in this gauge will be represented by the $\pi$ field, we will henceforth refer to this gauge as  {\it the  $\pi$ gauge}.  Let us also define $\hat{\pi}(x,r) \equiv -H \pi(x,r)$, so that 
 \be
 \tau(x) = \hat \pi(x, r_{UV})
 \ee  
 Solving for $N^i\equiv \pd_i \chi + N_T^i $ and $N\equiv 1+\delta N$ via the constraint equations $\sfrac{\delta S}{\delta N}=0$ and $\sfrac{\delta S}{\delta N^i}=0$ at linear order in $\pi$ gives
\begin{align}
\delta N_1=\vep H \pi\;, \quad 
\pd^2 \chi_1=-\frac{\vep}{ c_s^2}\frac{\pd}{\pd r}\left(H \pi\right)\;,
\quad N_{T,1}^i=0\;,
\end{align}
where $\vep$ and $c_s$ are given in equation (\ref{eq:EpsilonCs}). 
The bulk action \eqref{Bulk Action} in this gauge becomes
\begin{align}
S[\hat{\pi}]&=S_1[\hat{\pi}]+S_2[\hat{\pi}]+\dots\;,\nonumber\\
S_1[\hat{\pi}]&=\mplthr
\int\!\rmd r\, \rmd^2 x
\Bigg[
\frac{\pd }{\pd r}\left(-\frac{a(r)^2 \dot {H}\hat{\pi}}{H}\right)+a(r)^2 H \pd_i N_1^i
\Bigg]\;, \label{Spi1}\\
S_2[\hat{\pi}]&=
\frac{\mplthr}{2}\int\!\rmd r\,\rmd^2 x 
\Bigg[
\frac{a(r)^2\vep}{c_s^2}\left(\dot{\hat{\pi}}^2+\frac{c_s^2}{a(r)^2}(\pd\hat{\pi})^2
\right)
-\frac{\pd }{\pd r}\left(\frac{a(r)^2 \hat{\pi}^2  \ddot{H}}{H^2}\right)
\Bigg]
\;. \label{Spi2}
\end{align}
where the neglected terms in the first line come from higher powers of $Q$, and do not contribute to the quadratic action for $\hat \pi$.
Let us focus on the contribution to $S[\hat \pi]$ from the modes with small spatial momenta $k=|\vec k|$.  We provide a simplified but intuitive derivation of the $\tau$ action in this subsection, and leave a more rigorous version of this computation in appendix \ref{eq:RigorousActionComputations}. 
In the limit $k\to 0$, the EoM of $\hat \pi$ can be easily solved. Two linearly independent solutions are given by
\be
\hat \pi_{\rm cl}^{(1)}={\cal A}_1\int^r \!\rmd r'\, \frac{ c_s^2(r')}{a(r')^2\vep(r')}\;,\quad \hat \pi^{(2)}_{\rm cl}={\cal A}_2\;,
\ee
where ${\cal A}_1\;,{\cal A}_2$ are constant coefficients. The corrections from finite (but small) value of $k$ will appear at order ${\cal O}(k^2)$. Furthermore, as we show explicitly in Appendix \ref{eq:RigorousActionComputations}, the coefficient ${\cal A}_2$ must be equal to the boudary value for $\hat \pi(\vec k, r_{UV})=\tau(\vec k)$, while ${\cal A}_1= {\cal O}(k^2)$.

Therefore, to compute the quadratic on-shell action for $\hat \pi$ {\it at the  two derivative level} we simply set $\hat \pi( \vec k,r)={\hat \pi}^{(2)}_{\rm cl}=\tau(\vec k)$ and neglect the $\dot{\hat \pi}$ term.  We obtain
\begin{align}\label{zeta bdy action long}
S[{\hat \pi}_{\rm cl}]&\simeq \mplthr \int\!\rmd r\,\rmd^2 x \Bigg(\frac{\varepsilon(r)}{2}
(\pd \hat \pi)^2
\Bigg)\;\nonumber\\
&=\frac{\mplthr}{2} \Bigg(\int_{-\infty}^{+\infty}\!\rmd r\,\vep(r) \Bigg)
\int\!\frac{\rmd^2 k}{(2\pi)^2} k^2 \hat \tau(- \vec k)\hat \tau(\vec  k)\nonumber\\
&=\frac{\mplthr}{2} (L_{UV} - L_{\rm IR})
\int\!\frac{\rmd^2 k}{(2\pi)^2} k^2 \hat \tau(-\vec k)\hat \tau(\vec k)\;,
\end{align}
In the first equality, we have used the fact that the bulk geometry asymptotes to pure AdS, so we can safely drop the boundary terms in \eqref{Spi2} as well as terms that reside on the $x$-boundary;
in the third equality, we recalled the definition of $\vep$ in   \eqref{eq:EpsilonCs}, and then performed the $r$ integration explicitly by noting that $\vep(r) \rmd r=\rmd (1/H(r))$.  Note that this provides a sort of generalized junction condition relating the change in the UV and IR cosmological constants in the bulk and a kind of `integrated domain wall tension', even in the case where the `wall' is very thick in AdS units.

From holographic computations of the conformal anomalies \cite{Brown:1986nw, Henningson:1998gx, Skenderis:2002wp}, we know that the central charge of the CFT is related to the $AdS_3$ radius by
\be\label{Holo Anomaly}
c=12 \pi \mplthr L\;.
\ee
This means that we can write our result as
\begin{align}\label{pi bdy action}
S[\hat{\pi}_{\rm cl}]  
&=\frac{c_{\rm IR}-c_{\rm UV}}{24\pi}\int\!\frac{\rmd^2 k}{(2\pi)^2} 
k^2 \tau(-\vec k)\tau(\vec k) \;.
\end{align}
 This is precisely what we expect from \eqref{duality}
\be
-S_{\rm bulk}^{\rm onshell}[{\hat \pi}]= W_{\rm QFT}[\tau, \zeta=0] = S_{\rm WZ}[\tau]+\dots\;,
\ee
where dots denote nonlinear terms in $\tau$ and $S_{\rm WZ}[\tau]$ as given by \eqref{eq:2dGoldstoneWZAction}.
We have provided a holographic derivation of the Wess-Zumino action for $\tau$, reproducing the anomaly matching coefficient.
Note that the monotonicity of the c-function along the RG flow follows from the null energy condition (NEC) of the bulk action \cite{Myers:2010tj}. Indeed, to satisfying the NEC, we must demand $H(r)$ to be a monotonic decreasing function along the radial direction, $\dot{H}(r)<0$, so that
\be
c_{\rm UV}-c_{\rm IR}=12\pi \mplthr (L_{\rm UV}-L_{\rm IR})=12\pi \mplthr \int_{-\infty}^{+\infty}\frac{\rmd}{\rmd r} \left(\frac{1}{H}\right) \rmd r>0\;.
\ee
and so the central charge decreases under RG flow.

\subsection{UV and IR Conformal Anomalies from Holography}
\label{sec UV and IR Conformal Anomalies}

General conformal anomalies \cite{Deser:1993yx} were derived by Henningson and Skenderis \cite{Henningson:1998gx, Skenderis:2002wp} in a holographic context.  In an unperturbted CFT, their methods compute the conformal anomalies using only information about the region near a UV regulator surface.  We would like to obtain the conformal anomaly coefficients for both the UV and IR CFT using a unified approach, but clearly the anomalies of the IR CFT must depend on the bulk description far from the UV surface.  We will use methods very similar to our analysis in the previous section in order to give a unified treatment of UV and IR anomalies.

For this purpose we will let the 2d QFT live in a space with arbitrary metric.  The boundary metric can be written as
\be
h_{ij}^{(0)} = a(\rUV)^2 e^{2 \zeta_{(0)}({\vec x})}\delta_{ij}\;,
\ee  
where $\zeta_{(0)}({\vec x})$ encodes a Weyl factor.  We can compute correlators of $T^i_{\;\;i}$ in the CFT by varying with respect to $\zeta_{(0)}({\vec k})$.  By studying the large and small momentum behavior of these correlators, we obtain $c_{\rm UV}$ and $c_{\rm IR}$, respectively. 
For this purpose it is sufficient to set $\tau= 0$, or in other words, we can leave the conformal symmetry breaking couplings $\phi^{(0)}$ fixed.  As discussed above, $\zeta_{(0)}$ and $\tau$ are nearly equivalent, but they differ precisely in this context, where we wish to study the conformal anomaly.

With this choice of boundary conditions on the UV regulator surface, it is natural to let the bulk metric take the form
\be
h_{ij}=a(r)^2 e^{2 \zeta({\vec x},r)} \delta_{ij}\;,\quad \phi(\vec x, r)=\phi_{\rm bg}(r)\;,\quad \text{with }\;
\zeta({\vec x},\rUV)=\zeta_{(0)}(\vec x)\;,
\ee  
where the second condition above is equivalent to $\pi(\vec x, r)=0$.  We will call this gauge {\it the $\zeta$ gauge}, since the dynamical scalar degree of freedom in the bulk is the $\zeta$ field.  Our $\zeta$ is analogous to the notationally identical variable studied in cosmology \cite{Maldacena:2002vr, EFTInflation}.

As in the previous subsection, solving the constraint equations at linear order gives 
\begin{align}\label{constraints zeta gauge}
N&=1+\delta N\;, \quad N^i=\pd_i\chi+N_T^i \nonumber\\
\text{ with}\quad \delta N_1&=\frac{\dot{\zeta}}{H} \;,\quad \pd^2 \chi_1 =\frac{ \varepsilon}{c_s^2} \dot{\zeta}+\frac{\pd^2 \zeta}{a^2 H}\;,\quad N_{T,1}^i=0\;.
\end{align}
The full action \eqref{Bulk Action} in $\zeta$ gauge becomes
\begin{align}\label{zeta qua action}
S[\zeta]&\simeq \mplthr \int\!\rmd r\,\rmd^2 x \Bigg(\frac{a(r)^2 \varepsilon}{2c_s^2}\left[\dot{\zeta}^2+\frac{c_s^2}{a^2}(\pd \zeta)^2\right]
-\frac{\rmd }{\rmd r}\left[\frac{1}{2H}(\pd \zeta)^2\right]
\Bigg)\;.
\end{align}
Note that as promised, this action only differs from $S[\hat \pi]$ in equation (\ref{Spi2}) by a total derivative.
Varing this action, the equation of motion for $\zeta$ in the bulk reads
\be\label{EoM zeta}
\frac{\rmd }{\rmd r} \Bigg(\frac{a^2 \varepsilon}{c_s^2} \dot{\zeta}_{\rm cl}(\vec k,r)\Bigg)-\varepsilon k^2 \zeta_{\rm cl}(\vec k, r)=0\;,
\ee 
where $\zeta_{\rm cl}(\vec k, r)$ is the spatial fourier transform.  Since the $\hat \pi$ and $\zeta$ actions only differ by a boundary term, the $\zeta_{\rm cl}$ and $\hat \pi$ equations of motion are identical.

We cannot explicitly solve \eqref{EoM zeta} for general bulk configurations.  However, on physical grounds one would expect that it should be possible to determine the action for $\zeta$ at very large and very small momentum. 
Once again we only present a simplified version of the derivation here and leave a more rigorous computation to Appendix \ref{eq:RigorousActionComputations}. 

In the case of very large $k$, it is clear from equation (\ref{EoM zeta}) that if we neglect the variation of $a$, $\epsilon$, and $c_s$, then $\zeta_{cl}$ will decay exponentially as one approaches the IR, with a nonzero support deep within the UV region (with a width $\lesssim k^{-1}$).  
On the other hand, very close to the UV regulator surface, the geometry is approximately AdS and hence $\vep$ vanishes. 
This means that the action, when evaluated on the classical solution, will be entirely given by the total derivative term in equation (\ref{zeta qua action}), so with the aid of equation \eqref{Holo Anomaly} we find
\be
S[\zeta_{\rm cl}] \simeq
-\frac{c_{\rm UV}}{24\pi } \int_{k\gtrsim \Lambda_{\rm UV}}\!\frac{\rmd^2 k}{(2\pi)^2}\, k^2\zeta_{(0)}(-\vec k)\zeta_{(0)}(\vec k) 
\label{bdy action zeta high k}
\ee
in the limit of large $k$, where we are only probing the UV CFT.

In the case of very small $k$,  we can run a similar argument as in the previous subsection: the EoM of $\zeta$ \eqref{EoM zeta} can be easily solved, with the two linearly independent solutions given by
\be
\zeta_{\rm cl}^{(1)}={\cal A}_1\int^r \!\rmd r'\, \frac{ c_s^2(r')}{a(r')^2\vep(r')}\;,\quad \zeta^{(2)}_{\rm cl}={\cal A}_2\;,
\ee
where ${\cal A}_1\;,{\cal A}_2$ are constant coefficients satisfying ${\cal A}_2=\zeta_{(0)},\; {\cal A}_1=\zeta_{(0)} \times {\cal O}(k^2)$ (see Appendix \ref{eq:RigorousActionComputations} for a rigorous derivation).
To compute the quadratic on-shell action for $\zeta$ at {\it two derivative level } from \eqref{zeta qua action},  we simply set $\zeta(\vec k,r)=\zeta_{(0)}(\vec k)$ and neglect the $\dot{\zeta}$ term, so we get 
\begin{align} \label{zeta bdy action long}
S[\zeta_{\rm cl}] 
&\simeq \frac{\mplthr}{2} \Bigg(\int_{-\infty}^{+\infty}\!\rmd r\,\vep(r)-\frac{1}{H(r_{\rm UV})}\Bigg)
\int\!\frac{\rmd^2 k}{(2\pi)^2} k^2 \zeta_{(0)}(-\vec k)\zeta_{(0)}(\vec k)\nonumber\\
&=-\frac{\mplthr L_{\rm IR}}{2} 
\int\!\frac{\rmd^2 k}{(2\pi)^2} k^2 \zeta_{(0)}(-\vec k)\zeta_{(0)}(\vec k)\;,
\end{align}
Using the holographic relation from equation (\ref{Holo Anomaly}), we therefore find that
\begin{align}
S[\zeta_{\rm cl}]&\simeq
-\frac{c_{\rm IR}}{24\pi } \int_{k\lesssim \Lambda_{\rm IR}}\!\frac{\rmd^2 k}{(2\pi)^2}\, k^2\zeta_{(0)}(-\vec k)\zeta_{(0)}(\vec k)\;, \label{bdy action zeta low k}
\end{align}
where $c_{\rm IR}$ is the central charge for the IR boundary CFT.   The total derivative term in the bulk $\zeta$ action was crucial to obtain the factors of $c_{UV}$ and $c_{IR}$ at large and small momentum, differentiating the $\zeta_{(0)}$ action from the $\tau$ action.  Comparing equation \eqref{bdy action zeta high k} \eqref{bdy action zeta low k} with the expression for $W_{\rm QFT}$ given in Section \ref{sec:Flows}, we see immediately that $S_{\rm bulk}[\zeta_{\rm cl}]=-W_{\rm QFT}[\tau=0, \zeta_{(0)}]$, as was anticipated in \eqref{duality}.

%%%%%%%%%%%%%%%%%%%%%%%%%%%
%%%%%%%%%%%%%%%%%%%%%%%%%%%%%%
\subsection{Computations in (Asymptotic) Axial Gauge}

Is it possible to study a case in which both $\zeta_{(0)}$ and $\tau$ are present for the boundary generating function? To perform the corresponding holographic computation, we need to choose a new gauge, which is essentially the axial gauge. We will only summarize the main results in this subsection, leaving the detailed analysis to Appendix \ref{sec:AxialGauge}.

The axial gauge is defined by 
\be
N=1\;,\quad N_i=0\;,  \quad \phi(\vec x,r)=\phi_{\rm bg}(r+\pi)\;,\quad h_{ij}=a(r)^2 \Bigg[(1+2\zeta)\delta_{ij} +\pd_i \pd_j B \Bigg]\;.
\ee
For our purposes, axial gauge is not ideal because the conditions $N=1\;, N_i=0$ are not preserved by the Weyl transformations of equation (\ref{reduce Weyl}).  Nevertheless, there is a transformation that preserves the gauge and agrees with the equation (\ref{reduce Weyl}) to leading order in the fields.  Acting on the coordinates, we would have a transformation 
\be\label{true axial transf}
x^i\to x'^i=x^i-\pd_i\xi(\vec x) \int_{r_{\rm UV}}^r \frac{\rmd r_1}{a(r_1)^2}\;,\quad r\to r'=r+\xi(\vec x)\;.
\ee

An obvious disadvantage of the axial gauge preserving transformations is that they change the asymptotic behaviors of fields in the IR, due to the $r$-independence of $\xi$, so in order to get a regulated result from the bulk computation it is necessary to include an extra IR regulator brane.  Instead we can work in the approximate (or asymptotic) axial gauge\footnote{This is essentially the Fefferman-Granham gauge used in previous works, e.~g.~\cite{Henningson:1998gx}}, in which the $N=1\;, N_i=0$ conditions are satisfied only in the deep UV and IR region, but not in the intermediate region.  Residual transformations that preserve this choice are nothing but those in equation \eqref{true axial transf}, with $\xi$ promoted to be $r-$dependent and vanishing at deep IR. 

In the context of holography it is computationally complicated to treat gauge transformations as changes of coordinates, due to the existence of a fixed UV regulator brane, so we instead perform internal transformations induced from the spacetime gauge transformations, in which fields transform at each spacetime point, while we leave the coordinates unchanged. This internal transformation acts nonlinearly on the fields, and when restricted to the boundary, it coincides with the usual Weyl transformation, but only to leading order in fields.  The bulk action, constructed to be diff-invariant, will not be fully invariant under the internal transformations. Rather, the variation will yield precisely the trace anomaly for the boundary QFT. 

In Appendix \ref{sec:AxialGauge} we compute low-energy the holographic boundary action in this approximate axial gauge
\begin{align}\label{eq: axial gauge action}
S_{\rm axial}^{\rm on-shell}=\frac{c_{\rm UV}-c_{\rm IR}}{24\pi}\int\! \rmd^2 x
\Bigg((\pd \tau)^2+R^{(2)}\tau
\Bigg)
-\frac{c_{\rm IR}}{24\pi}\int\! \rmd^2 x\,(\pd \zeta_{(0)})^2+\dots\;,
\end{align}
which is what we expected from equation \eqref{duality}. However, it is worth mentioning an important caveat, namely, there are additional terms at the $\pd^2$ level (represented by ``$\dots$'') in the above expression. These terms are in fact invariant under the internal gauge transformations, and they would be absent if we had performed a true Weyl transformation \eqref{reduce Weyl}.

%%%%%%%%%%%%%%%%%%%%%%%%%%%
%%%%%%%%%%%%%%%%%%%%%%%%%%%%%%%%%%%
\section{Dilaton Actions in General Spacetime Dimension}
\label{sec:GeneralDimensions}

In this section, we will compute the low energy $\pi$ action in general (even) boundary dimension.   In $d\ge 4$, instead of using pure Einstein gravity, we will also include  a Gauss-Bonnet term in the action
\be\label{GB gravity}
S_{\rm Grav}=-\frac{M_{(d+1)}^{d-1}}{2}\int_{\cal M}
\sqrt{g}\bigg[R^{(d+1)}+\frac{\alpha}{(d-2)(d-3)}{\cal L}_{GB}
\bigg]-S_{GHY}\;,
\ee
where ${\cal L}_{GB}=R^2-4R_{\mu\nu}R^{\mu\nu}+R_{\rho\sigma\mu\nu}
R^{\rho\sigma\mu\nu}$, and $S_{GHY}$ denotes a suitable Gibson-Hawking-York boundary term, which would be required for a well-defined variational principle for this Gauss-Bonnet gravity action \cite{Wald:1984rg, Myers:1987yn}.  The motivation for including the Gauss-Bonnet term in the bulk theory is to distinguish the ``type A'' anomaly from the other conformal anomalies \cite{Deser:1993yx} in a general holographic calculation.  This permits a further consistency check of our methods.

We are mainly interested in the Wess-Zumino terms in the dilaton action \cite{Elvang:2012st, Elvang:2012yc}, which are responsible for the anomaly of the boundary QFT.  In a $d$ dimensional QFT these Wess-Zumino terms involve $d$ derivatives.   From a practical standpoint, it is difficult to compute the $\pd^d$ dilaton action holographically to nonlinear orders (in fields) in $d\ge 4$ boundary dimension. So, for the sake of simplification, some further assumptions about the structure of the action and the conformal breaking structure are needed:
\begin{itemize}
\item[({\it i})]  We focus on the matter action \eqref{Bulk Matter}, with $M_2=M_3=\dots =0$. This includes examples such as a minimally coupled scalar models with an action of the form 
\be\nonumber
S_m=\int\! \rmd^{d+1} x\sqrt{g}
\bigg(\!
-\frac{1}{2}(\pd\phi)^2-V(\phi)
\bigg)\;,\quad
\phi(x, r)=\phi_{\rm bg}(r+\pi(x,r ))\;.
\ee
\item[({\it ii})]   The solution interpolating between the UV and  IR geometry is very nearly AdS, i.e. in the region $r_2\le r \le r_1$, the variation of $H(r)$ over the radial direction is negligibly small. 
 \end{itemize}
Together, these two assumptions lead to a suppression of all terms in the $\pi$ (and hence $\tau$) action beyond the quadratic order in these fields.  However, we expect that in principle both could be relaxed in order to reproduce the full results of \cite{Elvang:2012yc}.
 
The second condition can be realized by specifying the form of $H(r)$ to be 
\be
H(r)=L_{\rm UV}^{-1}+\frac{f(r)}{\mpld^{d-1}L_{\rm UV}^d}\;,\quad |f(r)|\sim {\cal O}(1)\;,\quad f''\ll (f')^2\;,\dots, f^{(n)}\ll (f')^n\;,\dots\;,
\ee
where $H(r_1)\equiv L_{\rm UV}^{-1} \,,H(r_2)\equiv L_{\rm IR}^{-1}$.
This condition is an AdS analog for the slow-roll assumption in inflation.  We will also restrict to the regime 
\begin{align}\label{slow roll}
&\mpld\to \infty\;,\quad L_{\rm UV}\; \text{ fixed}\;,\quad \Delta L\equiv  L_{\rm UV}-L_{\rm IR}\propto\frac{1}{\mpld^{d-1}}\;.
\end{align}
The last relation was chosen so that $a_{\rm UV}^{(d)}\,, a_{\rm IR}^{(d)}\to \infty$, but the difference $a_{\rm UV}^{(d)}-a_{\rm IR}^{(d)}$ stays constant as $\mpld\to \infty$.  

As a consequence of these conditions, the gravitational action is completely `demixed' from the scalar mode $\pi.$\footnote{It was shown in \cite{EFTInflation} that the ``demix'' scale for the effective field theory of inflation is given by $E_{\rm demix}\sim H \epsilon^{1/2}$. Thus, in the parametric regime \eqref{slow roll}, it will be lower than any momenta/energies of the Fourier modes of $\pi$ under consideration.  }
In other words, after fixing to the $\pi$ gauge of Section \ref{sec:PiAction2d} and solving the constraint equations for $\delta N$ and $N^i$, one finds that $\delta N\,,N^i \propto \dot{H} \sim {\cal O}(\frac{\Delta L}{L_{\rm UV}})$. Therefore, the gravity action \eqref{GB gravity} and the mixing between $\pi$ and the metric will be subleading compared to terms in the Goldstone action.  Combined with assumption (i), which suppresses non-linear terms in the fields, this makes the computation of the full $\tau$ action very tractable.

The parameters $M_0$ and $M_1$ are determined via the background Einstein equation, as in the simpler case of pure Einstein gravity. They are given by 
\begin{align}\label{M0M1 GB}
M_0&=-(d-1)\mpld^{d-1}\left[\dot{H}+\frac{d}{2}H^2-\frac{1}{2}\alpha H^2 (d H^2+4\dot{H})\right]\nonumber\\
M_1&=-\frac{(d-1)\mpld^{d-1}}{2}\dot{H}\left(1-2\alpha H^2\right)\;.
\end{align}
Then the $\pi$  action following from \eqref{GB gravity} becomes
\be\label{Demix action GB}
S[{\hat \pi}]\simeq \frac{(d-1)\mpld^{d-1}}{2}\int\!\rmd r\, \rmd^d x \,a(r)^d \vep\left(1-2\alpha H^2\right)\bigg(\dot{\hat \pi}^2+\frac{(\pd {\hat\pi})^2}{a(r)^2}\bigg)+\dots\;,
\ee
where ${\hat \pi}=-H \pi$, $\vep\equiv -\dot{H}/H^2 \sim {\cal O}(\frac{\Delta L }{ L})$ and $\dots$ denotes terms of higher order in ${\cal O}(\frac{\Delta L }{ L})$. Remarkably we see that, at leading order, only the quadratic piece of the Goldstone action survives. Following from \eqref{Demix action GB}, the equation of motion for $\hat \pi$ in conformal radial coordinate $z$ (see Appendix \ref{eq:RigorousActionComputations} for detail) is
\be\label{EoM pihat demix}
\hat \pi''(z)-(d-1)a(z) H(z)\hat \pi'(z)-k^2 \hat \pi(z) =0\;,\quad z_1\le z \le z_2\;,
\ee
where $z$ is related to the $r$ coordinate by $\rmd z=-a(r)^{-1} \rmd r$.
Once again we neglect terms of higher order in ${\cal O}(\frac{\Delta L }{ L})$.  The EoM of $\hat{\pi}$ is only defined in the region $z_1\le z \le z_2$, because beyond this region the geometry is pure AdS and no spontaneous symmetry breaking occurs (or to put it another way,  the Goldstone action for $\pi$ vanishes identically since $\vep=0$ in the region $z\in [\zUV, z_1)\cup (z_2, +\infty)$). It is then natural to impose boundary conditions for $\hat{\pi}$ at $z=z_1$ and $z=z_2$ and require $\hat \pi$ to stay constant beyond these two points:
\be
\hat{\pi}(\vec k, z_1)=\tau(\vec k)\;,\quad \hat{\pi}(\vec k, z_2)=0\;. \label{BC pihat demix}
\ee

Notice that at the order we are working, the radial dependence of $a(z)$ and $H(z)$ can be fully determined:
\be
a(z)=\frac{L_{\rm UV}}{(1-\vep)z_{1}}\left(\frac{z}{z_{1}}\right)^{-1-\vep}\;,\;
H(z)=L_{\rm UV}^{-1}\left(\frac{z}{z_{1}}\right)^{\vep}\;.
\ee 
Using these relations  we can solve the EoM \eqref{EoM pihat demix} analytically. 
The solution subject to the boundary condition \eqref{BC pihat demix} is
\begin{align}
{\hat \pi}_{\rm cl}(\vec k, z)&=(k z)^\beta \bigg(
{\cal C}_1 K_\beta (k z)+{\cal C}_2 I_\beta (k z)
\bigg)\nonumber\\
&=\sum_{n=0}^\infty {\cal C}_1 w_{2n} (k z)^{2n}+{\cal C}_1 t_{2n} (kz)^{2n+2\beta}+{\cal C}_2 s_{2n} (kz)^{2n+2\beta}\;,\quad \text{ for } kz\ll1,
\end{align}
where 
\be
\beta\simeq \frac{d}{2}+\frac{d-1}{2}\vep\;, \quad {\cal C}_2=-{\cal C}_1 \frac{K_\beta (k z_2)}{I_\beta(k z_2)}\;,\quad {\cal C}_1 w_0=\tau(\vec k)\;,
\ee
and the various Taylor expansion coefficients are given by
\be
w_{2n}=\frac{2^{-2n-1+\beta} \pi \csc(\beta \pi)}{\Gamma(n-\beta+1)n!}\;,\quad t_{2n}=-\frac{2^{-2n-1+\beta}\pi \csc(\beta \pi)}{n!\Gamma(n+\beta+1)}\;,\quad s_{2n}=\frac{2^{-2n-\beta}}{n!\Gamma(n+\beta+1)}\;.
\ee
A key observation that will lead to a significant simplification is that
\begin{align}
&t_{2n} \sim {\cal O} \left(\frac{\Delta L }{ L} \right)\;,\quad s_{2n} \sim {\cal O}\left(\frac{\Delta L }{ L} \right)\;,\quad \text{ for all } n\in \mathbb{Z^+}\cup \{0\}\;,\nonumber\\
&a_{2n}\sim {\cal O}\left(\frac{\Delta L }{ L} \right) \text{ for } 2n \ge d\;,\quad a_{2n}\sim {\cal O}(1) \text{ for } 2n < d\;.
\end{align}
After some straightforward computations, we find that the low energy dilaton action (for modes with momenta $k \ll z_2^{-1}$) is given by
\be
S^{(d)}_{\rm bdy}&=&-\frac{(d-1)\mpld^{d-1}}{2}\int\! \frac{\rmd^d k}{(2\pi)^d} a(z_1)^{d-1} \vep\left(1-2\alpha L_{\rm UV}^{-2}\right)\hat{\pi}(\vec k, z_1)\hat{\pi}'(\vec k, z_1)\nonumber\\
&=&-\frac{d}{2}
\bigg(
\frac{(d-1)(1-2\alpha L_{\rm UV}^{-2})}{2^{d-1}\Gamma(d/2)\Gamma(d/2+1)}\left(\mpld L_{\rm UV}\right)^{d-1}\frac{\Delta L}{L_{\rm UV}}
\bigg)
\int\!\rmd^d x \,\tau(\vec x)\square^{d/2}\tau(\vec x) \nn \\
&=&-\frac{d}{2}\left(a_{\rm UV}^{(d)}-a_{\rm IR}^{(d)}\right)\int\!\rmd^d x \,\tau(\vec x)\square^{d/2}\tau(\vec x)\;,
\label{eq:DilationActionGeneralD}
\ee
where in the second equality we have used the fact that $H(z_1)/H(z_2)=L_{\rm IR}/L_{\rm UV}=(z_1/z_2)^\vep$, and in the last line we have used the formula for holographic $a$ anomaly (e.g. \cite{Myers:2010tj} in the conventions of \cite{Elvang:2012yc})  
\be\label{Holo a anomaly}
a^{(d)}=\frac{2^{-d+1}}{\Gamma(d/2)\Gamma(d/2+1)}\left(\mpld L_{\rm AdS}\right)^{d-1}
\bigg(1-\frac{2(d-1)}{d-3}\alpha L_{\rm UV}^{-2}
\bigg)
\;,
\ee
and expanded the difference $a_{\rm UV}^{(d)}-a_{\rm IR}^{(d)}$ to leading order in ${\cal O}(\frac{\Delta L}{L})$.  Once again, from the holographic picture, the monotonicity of the a-function --- $a^{(d)}_{\rm UV}-a^{(d)}_{\rm IR}>0$ --- is assured once the NEC is satisfied in the bulk.  Equation (\ref{eq:DilationActionGeneralD}) is the dilaton action in (even) $d$ dimensions modulo terms that vanish on shell.   Equivalent and more general results were obtained on the basis of symmetries in \cite{Elvang:2012yc}; our formula has been derived holographically, assuming a slow renormalization flow, as described above. 

%%%%%%%%%%%%%%%%%%%%%%%%%%%%%%%%%%%%%%%%%%%
%

\section{Higher-Derivative Operators and Multiple Fields}
\label{sec:Caveats}

In this section we will briefly consider more general bulk actions, including higher derivative interactions and multiple bulk fields.  The latter can be interpreted as RG flows involving several relevant operators added to the UV CFT action.  To study higher derivative terms we specialize to the 2d case, where we can include gravitational effects, and show that they do not affect our results concerning the anomaly matching.  Roughly speaking, this follows because the anomaly terms are determined by a matching procedure that only involves the equations of motion in the UV and IR region, where the conformal-breaking higher derivative terms vanish.  We study the presence of multiple fields in general dimensions, but in the demixed slow-flow limit of section \ref{sec:GeneralDimensions}, where it can be shown very easily that our results for conformal anomalies remain unchanged.  It would be interesting to understand how anomaly matching arises from unrestricted bulk dynamics.

\subsection{Higher Derivative Operators in the Bulk Action}\label{bulk action with K}
\label{sec:HigherDerivatives}

Thus far we have restricted our discussion to conformal-breaking matter actions of the form
\begin{equation*}
{\cal L}_M = \sqrt{g} \sum_{n=0}^{\infty} \frac{1}{n!}M_{n}(r+\pi)Q^n \;.
\end{equation*}
where $Q$ was defined in equation (\ref{eq:QDef}).
What we will show in this subsection is that at least for 2d QFTs, the correspondence between the bulk effective action and a boundary anomaly matching holds in a more general context.

In particular, there are other terms beyond $Q$ that are invariant under the {\it spatial} diffeomorphisms  $x^i\to x^i+\xi^i(x,r)$ and the diagonal $r$ diffeomorphism $r \to r + \sigma(x, r),\; \pi\to \pi -\sigma (x,r)$.  Therefore these terms should also be included in the bulk effective action.  Up to fourth order in  derivatives there are three new terms, constructed from the extrinsic curvature of constant $r$ slices, $K_{ij}$: 
\begin{align}
\Delta {\cal L}_1 &=f_1(r+\pi) N \sqrt{h}(N-1) \delta E^i_{\;\;i} 
\;, \\
\Delta {\cal L}_2 &=f_2(r+\pi) N \sqrt{h} (\delta E^i_{\;\;i})^2
\;, \\
\Delta {\cal L}_3 &=f_3(r+\pi) N \sqrt{h} (\delta E^i_{\;\;j} \delta E^i_{\;\;j})\;,
\end{align}
where $E_{ij}$ is related to the extrinsic curvature by $E_{ij}=N K_{ij}$, and 
\be
\delta E_{ij}\equiv E_{ij}+H h_{ij}\;.
\ee
Including these terms, the bulk matter action becomes 
\begin{align}\label{high k correction to bulk action}
S_M&=\int\!\rmd r \rmd^2 x\sqrt{g} \sum_{n=0}^{\infty} \frac{1}{n!}M_{n}(r+\pi)Q^n 
+\int\!\rmd r \rmd^2 x\Big(\Delta {\cal L}_1+\Delta {\cal L}_2+\Delta {\cal L}_3\Big)\;.
\end{align}
We choose to work in the $\zeta$ gauge; as we will see shortly these new higher derivative terms will not change the results we obtained in the  previous subsection \ref{sec UV and IR Conformal Anomalies}: the on-shell bulk action will still be given by \eqref{bdy action zeta high k} and \eqref{bdy action zeta low k} at quadratic order in $\zeta$.

 Once again we need to solve the constraint equations for $\sfrac{\delta S}{\delta N}=0$ and $\sfrac{\delta S}{\delta N^i}=0$ to linear order. Schematically the solution takes the form of 
\be
\delta N_1=r_1 \dot{\zeta} +r_2 \pd^2\zeta\;,\quad \pd_i N_1^i=t_1 \dot{\zeta}+t_2 \pd^2 \zeta\;,\quad N_{1,T}^i=0\;.
\ee
These coefficients $r_1\,, r_2\,,t_1\,,t_2$ are functions of $f_1\,, f_2\,, f_3\,, H\,, \dot{H}\,$ and $\mplthr$.  In the limit $f_1=f_2=f_3=0$, the above expressions for $\delta N_1$ and $N_1^i$ reduce to \eqref{constraints zeta gauge}.

Expanding the new bulk action to quadratic order in $\zeta$, we have
\begin{align}\label{full bulk action}
\tilde{S}_{\rm bulk}^{(2)}&= \mplthr \int\!\rmd r\,\rmd^2 x \Bigg(\frac{a(r)^2 \varepsilon}{2c_s^2}\left[\dot{\zeta}^2+\frac{c_s^2}{a^2}(\pd \zeta)^2\right]
-\frac{\rmd }{\rmd r}\left[\frac{1}{2H}(\pd \zeta)^2\right]
\Bigg)+\Delta S_M\;,
\end{align}
with 
\begin{align}
\Delta S_M&=\int\!\rmd r\,\rmd^2 x\,a^2 \left(f_1 \delta N_1(\pd_i N_1^i-2\dot\zeta)+f_2(\pd_i N_1^i-2\dot\zeta)^2+f_3(\pd_i N_1^i)^2\right)\nonumber\\
&=\int\!\rmd r\,\rmd^2 x \left(w_1(r) \dot{\zeta}^2+w_2(r) \dot{\zeta} \pd^2 \zeta +w_3(r)(\pd^2\zeta)^2\right)\nonumber\\
&=\int\!\rmd r\,\rmd^2 x \left[w_1(r) \dot{\zeta}^2+\frac{\dot{w}_2}{2} (\pd \zeta)^2 +w_3(r)(\pd^2\zeta)^2-\frac{\rmd}{\rmd r}\left(\frac{w_2}{2}(\pd\zeta)^2\right)\right]\;,
\end{align}
where in the last equality we have integrated by parts. 
The explicit expressions for $w$'s are lengthy and not very useful; they vanish when $f_1=f_2=f_3=0$.

Notice that the parameters $f_1\,, f_2\,, f_3\,$ must vanish in the UV (say, $r>r_1$, with $r_1$ the scale beyond which the geometry is pure AdS to sufficient precisions) and the IR ($r<r_2$) region, for the same reason that $M_2$ vanishes (roughly speaking, we demand that the geometry be purely AdS and that background fields $\phi_{\rm bg}$ approach a constant in the deep UV and IR region,  so there will be no breaking of $r$ translations or any other breaking of the AdS isometries, 
see Appendix \ref{eq:RigorousActionComputations} for a detailed discussion).   As an immediate consequence, in the UV and IR region, $w_1=w_2=w_3=0$, and the bulk action is unaltered under the inclusion of the higher derivative operators. 

In the small momentum limit $k\to 0$, the EoM for $\zeta$ following from the new bulk action \eqref{full bulk action} becomes
\be
\frac{\rmd }{\rmd r}\left(\left(\frac{a(r)^2\vep(r)}{2c_s^2(r)}+w_1(r)\right)\dot{\zeta}_{\rm cl}(\vec k, r)\right)=0\;.
\ee  
The general solution of the above equation is given by
\be
\lim_{k\to 0}\zeta(\vec k, z)= {\cal A}_2+{\cal A}_1 \int\!\rmd r \left(\frac{a(r)^2\vep(r)}{2c_s^2(r)}+w_1(r)\right)^{-1}\;,
\ee
which has an obvious resemblence with our prior results (see Eqn.~\eqref{zeta sol small k} for details). As before, we have  ${\cal A}_2=\zeta_{(0)},\; {\cal A}_1=\zeta_{(0)} \times {\cal O}(k^2)$.

We can now repeat the analysis in Section \ref{sec UV and IR Conformal Anomalies} to compute contribution to the on-shell bulk action at two derivative level; we find that the higher derivative terms $\Delta S_M$ will not contribute to the low energy boundary $\zeta$ action,
\begin{align}
\Delta S_M\simeq \int\!\rmd^2 x (\pd\zeta_{(0)})^2\left(\int_{-\infty}^{+\infty}\rmd r\frac{\dot{w}_2}{2}\right)-\int_{r=\rUV}\!\rmd^2 x\left(\frac{w_2}{2}(\pd\zeta)^2\right)=0\;,
\end{align}
since $w_2$ vanishes in the UV and IR region. As for high $k$ modes, the analysis is even simpler:  as we argued before, new terms in the bulk action \eqref{high k correction to bulk action} will not change its form in the UV region ($r>r_1$), and only the neighborhood of the UV regulator surface controls the behavior of UV CFT modes, so once again we are led to the same conclusion.  Even in the presence of higher derivative operators in the effective bulk action, the anomaly terms in the on-shell action for $\zeta_{(0)}$ are still given by \eqref{bdy action zeta high k} and \eqref{bdy action zeta low k}.  Similar results hold for the $\tau$ action computed from the bulk $\pi$ field.

\subsection{Multiple Bulk Fields}
\label{Multi-fields}

We would like to know if we need to make any assumptions about the number of degrees of freedom in the bulk. One would naively expect that other bulk fields could contaminate anomaly matching through their interactions with $\pi$.  In this section we will study this question for the case of QFTs in general boundary dimension $d$, but we will restrict ourselves to studying the quadratic action in the ``demix'' regime of section \eqref{slow roll}.  Thus the mixing between the matter fields and gravity will be subdominant, and it will be sufficient to treat the spacetime geometry as a fixed background.  

Furthermore, given the difficulty of studying the most general models, we will focus on the case where the $N$ scalars $\sigma_I$ are Goldstone fields of a spontaneously broken $U(1)^N$ symmetry.   Because of the shift symmetry, the $\sigma_I$ will be derivatively coupled. These assumptions are sensible since they imply that the $\sigma_I$ are dual to marginal operators; our results would not change if we included a small negative squared masses for these fields, so that they would be dual to relevant operators in the UV CFT.

We can adapt results from a study of the EFT of Inflation in the presence of multiple fields \cite{Senatore:2010wk} in order to write the quadratic Lagrangian as
\be\label{multifield bulk action}
S_{\rm multifield}=\frac{\mpld^{d-1}}{2}\int \rmd r \rmd^d x \,a^d(r) \Bigg[
-\frac{\dot H}{c_s^2} \left(  \dot \pi^2 + \frac{c_s^2}{a^2(r)} (\partial \pi)^2 \right) + 2{\tilde M}_1^{I} \dot \pi \dot \sigma_I + \mathcal{L}(\sigma_I)   
 \Bigg]\;.
\ee
We assume that the kinetic terms in ${\cal L}(\sigma_I)$ have order unity coefficients, and that the other  scalar fields $\sigma_I$ have a boundary condition $\sigma_I = 0$ on the UV boundary.  

Due to the kinetic mixing, $\pi$ behaves like an external source in the EoM for $\sigma_I$, and vice versa.   The mixing is important when the mixing strength $\tilde{M}_1^I$ is roughly of the order $|\dot{H}/c_s^2|^{1/2}$. We will henceforth assume that $\tilde{M}_1^I\sim |\dot{H}/c_s^2|^{1/2}$
\footnote{When $\tilde{M}_1^I$ is far away from this central value, it will lead to ghosts or negligible mixing \cite{Senatore:2010wk}.}
,
and that all the parameters in the action \eqref{multifield bulk action}, such as $H\,, c_s^2\,, \tilde{M}_1^I\,\dots$, have very weak $r-$ dependence (i.e.~ of order $\Delta L/L$ suppressed). 

In contemplating the EoM for $\sigma_I$, which follows from \eqref{multifield bulk action},
we find that the classical solution for $\sigma_{I}$ consistent with the boundary conditions can have non-zero supports $\sim \CO\left(\frac{\Delta L}{L}\right)$ only within $r_2\le r\le r_1$, since the source for its EoM, $\sim \frac{\rmd }{\rmd r}\left(a^d \tilde{M}_1^I \dot\pi\right)\sim \CO\left(\frac{\Delta L}{L}\right)$, vanishes in the UV  and the IR region. Therefore, evaluated on the classical solutions, the multifield bulk action \eqref{multifield bulk action} at leading order in slow flow becomes
\be\label{multifield bulk action OS}
S_{\rm multifield}^{\rm on-shell}\simeq\frac{\mpld^{d-1}}{2}\int\! \rmd r \,\rmd^d x \, \frac{\rmd}{\rmd r}\Bigg(\frac{a^d (r)\vep}{c_s^2}\dot{\hat{ \pi}}_{\rm cl}\hat {\pi}_{\rm cl} \Bigg)\;,
\ee
where we have used the fact that both $\sigma_I$ and $\dot{\sigma}$ vanishes on the UV boundary.

In return, the negligibly small $\sigma^I$, as a source in the EoM of $\pi$, can not change significantly the classical solution of $\pi_{\rm cl}$. This can also be verified by running the same ``matching'' procedure in Appendix  in Section \ref{eq:RigorousActionComputations}.  
Therefore we conclude that \eqref{multifield bulk action OS} will yield the same on-shell action at order $\pd^d$ as \eqref{eq:DilationActionGeneralD} \eqref{Holo a anomaly}, with $\alpha=0$.

    It is not obvious whether we can neglect the $\sigma_I$ in higher dimensions, especially if we wish to compute the bulk $\pi$ action to nonlinear order; however this should be the case due to the universality of the $a$-anomaly.  To investigate this further it would likely be useful to understand the relationship between the symmetry constraints on the spurion (dilaton) action in the CFT and the constraints of diffeomorphism-invariance on the $\pi$ action, since ultimately these must play the same role in guaranteeing a specific form for the action evaluated on $\pi_{UV}$.   

\section{Discussion}
\label{sec:Discussion}

We have shown that in holographic descriptions of 2d QFTs, and higher dimensional QFTs with a slow renormalization flow, one can derive the A-type anomaly coefficient from a universal matching procedure in the bulk.  In the 2d case we also illustrated anomaly matching by showing how the computation of the dilaton action relates to the computation of the anomalies of the UV and IR CFTs.  It would be very interesting to understand the universality of anomaloy matching in complete generality, without the slow-flow assumption.  It might also be interesting to study the constraints from AdS/CFT Ward identities \cite{Berezhiani:2013ewa, Berezhiani:2014tda, Hinterbichler:2013dpa, Goldberger:2013rsa} related to the consistency relation \cite{Maldacena:2002vr, EFTInflation, Cheung:2007sv, Creminelli:2011mw, Creminelli:2012ed} in cosmology.  The effective theory we develop may have broader applications for the study of holographic phenomenology \cite{Goldberger:1999uk, Chacko:2013dra}, where our methods could be used to derive both the universal anomaly matching terms and the conformally invariant terms for light dilatons.  Our methods might also relate to the  study of holographic entropy \cite{Bousso:2002ju}, particularly insofar as the $\tau$ action has recently been tied \cite{Banerjee:2014daa} to computations of entanglement entropy \cite{RT1, RT2}.  One might also study boundary \cite{Cardy:1986gw, Cardy:2004hm, Liendo:2012hy} and interface CFTs, or even RG domain walls \cite{Gaiotto:2012np} via holography, generating a bulk domain wall and a $\pi$ and $\tau$ action in only half of the spacetime.  In that case the $\pi$ field might interpolate between the dilaton and the displacement operator in the ICFT.

\section*{Acknowledgments}

We thank Cyrus Faroughy and Liam Fitzpatrick for discussions and comments on the draft.  This work was supported in part by the National Science Foundation grant PHY-1316665.

%%%%%%%%%%%%%%%%%%%%%%%%%%%%%%%%%%%%%%%%%%
%%%%%%%%%%%%%%%%%%%%%%%%%%%%%%%%%%%%%%%%%%%%
\appendix

\section{Complete Computations of $\pi$ and $\zeta$ Actions}
\label{eq:RigorousActionComputations}

In this section, we provide a rigorous derivation of equation \eqref{pi bdy action}, \eqref{bdy action zeta high k} and \eqref{bdy action zeta low k}. We will focus our discussion mainly on the holographic computations in the $\zeta$ gauge. But the same result can be used to facilitate the computations in the $\pi$ gauge.  

Recall that the background bulk metric under consideration asymptotes to pure AdS in the deep UV and IR region (Eqn.~\eqref{IR UV Behavior}), but can take an arbitrary form (with $H(r)>0,\; \dot{H}(r)<0$) in between. 
To be quantitatively specific, let us assume that $H(r)$ and background scalar value $\phi_{\rm bg}(r)$ vary only within the interval $r\in [r_2,r_1]$ (the domain wall regime), and they remains constant in the deep UV region $r> r_1$ and the deep IR region $r< r_2$: 
\begin{align}
H(r>r_1)=L_{\rm UV}^{-1}\;,\quad  H(r<r_2)=L_{\rm IR}^{-1}\;.
\end{align}  
For a given bulk matter action with a potential, this requirement is equivalent to demanding the $\phi_{\rm bg}$ field rests at the local minima of the potential in the deep UV and deep IR region, so that the potential energy contributes effectively as a cosmological constant. In our language of effective field theory,  we must have 
\be
M_2=M_3=\dots =0\;,\quad \text{for }   r\in (-\infty, r_2) \cup (r_1, +\infty)\;,
\ee
since $\dot{\phi}_{\rm bg}=0$ there.  

Note that in principle the EoM for $\zeta$ field, following from the quadratic action \eqref{zeta qua action}, cannot be defined outside the domain wall regime, since $\vep$ vanishes there.  
To simplify the computation, we employ the following trick, by first assuming $H(r)$ still varies {\it slowly} in the near UV and near IR region, and then taking this $r-$dependence to zero.   That is, instead of treating $H$ as a constant, we assume that 
\be
\varepsilon=\varepsilon_0={\rm constant}\;,\quad \text{for } r\in (-\infty, r_2) \cup (r_1, +\infty)\;.
\ee
The EoM for $\zeta$, given by \eqref{EoM zeta} is then valid over the whole bulk space.

\subsection{Conformal Radial Coordinate}
It turns out convenient to choose another radial coordinate $z(r)$, defined by 
\be\label{def z coo}
\rmd z =- \frac{1}{a(r)} \rmd r = -e^{-A(r)} \rmd r \;.
\ee
Correspondingly the domain wall spreads from $z_1\equiv z(r_1)$ to $z_2\equiv z(r_2)$.
Noting that for any scalar function of $r$
\be
\dot{f}(r)=-\frac{1}{a(z)}\frac{\rmd}{\rmd z}f(z)\equiv -\frac{1}{a(z)}f'(z)\;,
\ee
The EoM of $\zeta$ in $z$ coordinate becomes 
\be\label{EoM zeta in z}
\frac{\rmd }{\rmd z} \Bigg(\frac{a(z) \varepsilon(z)}{c_s^2(z)} \zeta_{\rm cl}'(\vec k,z)\Bigg)-a(z)\varepsilon(z) k^2 \zeta_{\rm cl}(\vec k, z)=0\;,
\ee 
where prime $(')$ denotes derivative with respect to $z$.

In the deep UV and IR region, the background bulk metric has a simple $r-$dependence that can be analytically solved:
\begin{align}\label{a H in UV}
a(z)=\frac{1}{(1-\vep_0)z_{1}H_{1}}\left(\frac{z}{z_{1}}\right)^{-\sfrac{1}{1-\vep_0}}\;,\;
H(z)=H_{1}\left(\frac{z}{z_{1}}\right)^{\sfrac{\vep_0}{1-\vep_0}}\;,\; 
c_s(z)=1\;,\;\text{for } \zUV\le z \le z_1\;, 
\end{align}
where $\zUV=z(r_{\rm UV})\,,\; H_{1}=H(r_{1})=L_{\rm UV}^{-1}$. We also have similar expressions for $a(z)$ and $H(z)$ in the deep IR region, $z\ge z_2$.  

\subsection{Solving EoM for $\zeta$}
By knowing the radial dependence,  we can solve analytically the EoM \eqref{EoM zeta in z} in the region $ z\in [\zUV, z_1) \cup (z_2, +\infty)$. The general solution that is regular in large $z$ reads
\begin{align}
\zeta_{\rm cl}(\vec k, z)&=(k z)^{\beta}
	\left[
		{\cal C}_1(\vec k) I_{\beta}( k z)+{\cal C}_2(\vec k) K_{\beta}( k z)
	\right]
	\;,\quad \text{for } \zUV\le z \le z_1\;,\label{zeta sol UV}\\
\zeta_{\rm cl}(\vec k, z)&=(k z)^\beta {\cal D}(\vec k)   K_\beta (k z)
	\;,\quad \text{for }  z \ge z_2\;,
	 \label{zeta sol IR}
\end{align}
where $I_\beta$ and $K_\beta$ are modified Bessel functions of the first and second kind, and $\beta$ is given by 
\be
\beta=\frac{2-\vep_0}{2(1-\vep_0)}\simeq 1+{\cal O}(\vep_0)\;.
\ee

We are about to determine the coefficients ${\cal C}_1,\;{\cal C}_2$, and ${\cal D}$ as functions of $\vec k$.  When the momentum of the mode function $k=|\vec k|$ is large, i.~e.~$ k z_1\gg 1$, it is easy: in order that $\zeta$ is regular, we must have 
\be\label{coeff high k}
{\cal C}_1(\vec k)=0\;,\quad {\cal C}_2(\vec k)=\zeta_{(0)}(\vec k)\;,\quad \text{for } kz_1\gg 1\;,
\ee
and the coefficient ${\cal D}$ is not important, since $K_\beta(x)\sim x^{-1/2}e^{-x}$, and hence $\zeta_{\rm cl}$ is highly suppressed in the region $z\ge z_2>z_1\gg k^{-1}$.

For the opposite parametric region with small momenta, $k z_2 \ll1 $, the computaion is much more involved. We use the ``matching'' procedure to relate the UV behavior of $\zeta_{\rm cl}$ with its IR behavior,  which will be explained in detail now. 
By Taylor expanding $I_\beta (k z)$ and $K_\beta (k z)$ in powers of $k z$, we get
\begin{align}
\zeta_{\rm cl}(\vec k, z)&= (k z)^{2\beta}( {\cal C}_1 a_0+{\cal C}_2 \tilde{b}_0 \dots)+{\cal C}_2(b_0+b_2 (k z)^2+\dots)\;,\quad \text{for } \zUV\le z \le z_1 \text{ and } k z \ll1 \;, \label{zeta  UV series}\\
\zeta_{\rm cl}(\vec k, z)&= (k z)^{2\beta}({\cal D} \tilde{b}_0+\dots)+{\cal D}(b_0+b_2 (k z)^2+\dots)\;,\quad \text{for }z\ge z_2 \text{ and } k z \ll1 \;, \label{zeta IR series}
\end{align}
with the coefficients given by 
\begin{align}\label{Taylor coeff}
a_0&=\frac{2^{-\beta }}{\Gamma(\beta +1)}\;,\quad
b_0=2^{\beta-1}\Gamma(\beta)\;,\quad 
b_2=\frac{2^{\beta -3} \Gamma (\beta )}{1-\beta }\;,\quad
\tilde{b}_0=2^{-\beta-1}\Gamma(-\beta)\;.
\end{align}

On the other hand, in the region $\zUV \le z\lesssim k^{-1}$,  we solve \eqref{EoM zeta in z} perturbatively in $k=|\vec k|$.  At the zeroth order ($k\to 0$ limit), the solution reads
\begin{align}\label{zeta sol small k}
\zeta_{\rm cl}(\vec k, z)\simeq{\cal A}_2+{\cal A}_1\int_{z_c}^z \!\rmd z'\, \frac{ c_s^2(z')}{a(z')\vep(z')}\;,\quad \text{for } z\in [\zUV, \infty) \text{ with } k z \ll 1\;,
\end{align}
where $z_c$ can be any arbitrary reference point. 

This intermediate, low $k$ solution \eqref{zeta sol small k} must match the solutions in deep UV region and deep IR region, Eqn.~\eqref{zeta sol UV} and Eqn.~\eqref{zeta sol IR}, respectively, in their overlapping domain.  Consider the modes with low momentum $k$ with $k\ll z_2^{-1} < z_1^{-1}$. The matching requires that 
\begin{align}
{\cal C}_2 b_0 &= {\cal A}_2+{\cal A}_1\int_{z_c}^{z_1} \!\rmd z'\, \frac{ c_s^2(z')}{a(z')\epsilon(z')}-Q_{1}{\cal A}_1z_1^{2\beta}\;,\quad 
{\cal C}_1 a_0+{\cal C}_2 \tilde{b}_0 = k^{-2\beta}{\cal A}_1 Q_{1}\;, \\
{\cal D} b_0 &= {\cal A}_2+{\cal A}_1\int_{z_c}^{z_2} \!\rmd z'\, \frac{ c_s^2(z')}{a(z')\epsilon(z')}-Q_{2}{\cal A}_1z_2^{2\beta}\;,\quad 
{\cal D} \tilde{b}_0 = k^{-2\beta}{\cal A}_1 Q_{2}\;,
\end{align}
where 
\be\label{Q}
Q_i = \frac{(1-\vep_0)^{2}}{\vep_0(2-\vep_0) } z_i^{-\sfrac{\vep_0}{1-\vep_0}}H_{i}\;,\quad i=1,2\;.
\ee
Eliminating ${\cal A}_1$ and ${\cal A}_2$ from the above equations, we get relations between the UV and IR coefficients: 
\be\label{coeff low k}
{\cal C}_2 = {\cal D}+{\cal O}(k^{2\beta})\;,\quad 
{\cal C}_1 a_0={\cal C}_2 \tilde{b}_0 \left(\frac{Q_{1}}{Q_{2}}-1\right)+{\cal O}(k^{2\beta})\;,
\quad \text{for } kz_2\ll 1\;.  
\ee

\subsection{Induced Boundary Action in $\zeta$ Gauge}
Now we are in the position to compute the onshell bulk action. For the high $k$ regime, $k\gtrsim z_1^{-1}$, as we showed in previous subsection, $\zeta_{\rm cl}(\vec k, z)$ exponentially decays and hence is significantly nonzero only in the vicinity of the UV boundary, $z\lesssim k^{-1}\lesssim z_1$. Therefore only the (UV) boundary term in \eqref{zeta qua action} contributes to the onshell action.  With the aid of \eqref{coeff high k}, we have  
\begin{align}\label{high k zeta rigo}
\lim_{\zUV\to 0}\lim_{\vep_0 \to 0} S[\zeta_{\rm cl}] &\simeq \mplthr
\int_{k\gtrsim z_1^{-1}}\!\,\frac{\rmd^2 k}{(2\pi)^2} \Bigg(
-\frac{k^2}{2H(\zUV)} \zeta(-\vec k, \zUV)\zeta(\vec k,\zUV)
\Bigg)\nonumber\\
&=-\frac{\mplthr L_{\rm UV}}{2} 
\int_{k\gtrsim z_1^{-1}}\!\,\frac{\rmd^2 k}{(2\pi)^2} 
k^2 \zeta_{(0)}(-\vec k)\zeta_{(0)}(\vec k)\;.
\end{align}
Notice that the limit $\vep_0\to 0$ should be taken {\it before} the $\zUV\to 0$. 

While in the low $k$ regime, $k\lesssim z_2^{-1}$, using \eqref{a H in UV}, \eqref{zeta  UV series},  \eqref{Taylor coeff}, \eqref{Q} and \eqref{coeff low k},  the $\zeta$ action \eqref{zeta qua action}  becomes 
\begin{align}\label{low k zeta rigo}
\lim_{\zUV\to 0}\lim_{\vep_0 \to 0} S[\zeta_{\rm cl}] &\simeq \mplthr
\int\!\,\frac{\rmd^2 k}{(2\pi)^2} \Bigg(
-\frac{a(z) \varepsilon_0}{2}\zeta(-\vec k,\zUV)\zeta'(\vec k, \zUV)
-\frac{k^2}{2H(\zUV)} \zeta(-\vec k, \zUV)\zeta(\vec k,\zUV)
\Bigg)\nonumber\\
&=-\frac{\mplthr L_{\rm IR}}{2} 
\int_{k\lesssim z_2^{-1}}\!\,\frac{\rmd^2 k}{(2\pi)^2} 
k^2 \zeta_{(0)}(-\vec k)\zeta_{(0)}(\vec k)\;,
\end{align}
where we have used the fact that $\zeta_{(0)}(\vec k)\equiv {\cal C}_2 b_0$ is understood as the metric on the boundary.  The equation \eqref{high k zeta rigo} and \eqref{low k zeta rigo} match what we got in Section \ref{sec UV and IR Conformal Anomalies} from a simplified version of the derivation. 

\subsection{Induced Boundary Action in the $\pi$ Gauge}
In this subsection, we are going to compute the on-shell bulk action in $\pi$ gauge, using the ``matching'' method developed in previous subsections of this appendix.  Regarding the absence of the dilaton field $\tau$ in high energy scale, we focus on computing the onshell action for low momenta, $k\lesssim z_2$.  

Noting that the EoM for $\hat{\pi}$ in $z$ coordinate, following from the quadratic action \eqref{Spi2}, is identical to equation \eqref{EoM zeta in z},
we see immediate that the low momentum solution of $\hat{\pi}_{\rm cl}$ also takes the form of \eqref{zeta IR series}, with the Taylor series coefficients $a_n$'s and $b_n$'s given by equation \eqref{Taylor coeff} and $\CC_1, \CC_2$ by equation \eqref{coeff low k} and \eqref{Q}.

Plugging the classical solution of $\hat{\pi}_{\rm cl}$ into equation \eqref{Spi2},  dropping terms with $\ddot{H}\propto \vep_0^2$, and then performing the same computation as in the case of low momenta $\zeta$ action provided in the last subsection,we obtain that 
\begin{align}\label{pi bdy action rigo}
\lim_{\zUV\to 0}\lim_{\vep_0 \to 0}S[\hat{\pi}_{\rm cl}] &\simeq \mplthr
\int\!\,\frac{\rmd^2 k}{(2\pi)^2} \Bigg(
-\frac{a(z) \varepsilon_0}{2}\hat{\pi}(-\vec k,\zUV)\hat{\pi}'(\vec k, \zUV)
\Bigg)\nonumber\\
&\simeq \frac{\mplthr}{2}\big(L_{\rm UV}-L_{\rm IR}\big)
\int\!\frac{\rmd^2 k}{(2\pi)^2} 
k^2 \tau(-\vec k)\tau(\vec k)\nonumber\\
&=\frac{c_{\rm UV}-c_{\rm IR}}{24\pi}\int\!\frac{\rmd^2 k}{(2\pi)^2} 
k^2 \tau(-\vec k)\tau(\vec k)\;,
\end{align}
where in the last step, we have used \eqref{Holo Anomaly}. It matches equation \eqref{pi bdy action}, confirming our intuitive derivation given in Section \ref{sec:PiAction2d}

%%%%%%%%%%%%%%%%%%%%%%%%%%%
%%%%%%%%%%%%%%%%%%%%%%%%%%%%%%%%%%

\section{The (Asymptotic) Axial Gauge for $d=2$}
\label{sec:AxialGauge}

One may wonder if one can choose arbitrary combinations of boundary values for $g_{(0)}$ and $\tau$. Obviously neither the $\pi$ gauge nor the $\zeta$ gauge is appropriate in studying this question. So in this section, we will redo the previous computation of the bulk on-shell action in a different gauge, namely the {\it axial gauge}, in which  
we use the bulk gauge freedom to set $N=1\;, N_i=0$ identically. The matter field, as in the $\pi$ gauge, is written as $\phi(\vec x,r)=\phi_{\rm bg}(r+\pi)$, 
and the linearized spatial metric as
\be
h_{ij}=a(r)^2 \Bigg[(1+A)\delta_{ij} +\pd_i \pd_j B \Bigg]\;,
\ee
where we once again neglected the vector perturbations.  

We can work out the (linearly independent) equations of motion in this gauge. At linear order in perturbations, they are given by 
\begin{align}
&\frac{H}{2}\dot{\psi}+\frac{\pd^2 A}{2a(r)^2}+\frac{\dot{H}\dot{\pi}}{c_s^2}=0\;,\label{pidot eq}\\
&\dot{A}=2\dot{H}\pi\;,\label{Adot eq}\\ 
&\ddot{\psi}+2 H \dot{\psi} =0\;, \label{EoM psi}
\end{align}
where $\psi\equiv \pd^2 B$. 
The EoM for $\psi$ is decoupled, and hence can be solved independently.  The general solution reads
\be\label{psi sol}
\psi_{\rm cl}=\psi_{(0)}(\vec x)+2{\cal F}(\vec x)\int_{r_{\rm UV}}^r\frac{\rmd r'}{a(r')^2}\;.
\ee 
For $\CF\ne 0$, the solution $\psi$ diverges in the IR ($r\to -\infty$). Now we solve the coupled equations of motions for $A$ and $\pi$.
Plugging \eqref{psi sol} and \eqref{Adot eq} into \eqref{pidot eq}, we have 
\be\label{EoM v}
\frac{\rmd }{\rmd r}\left(\frac{a(r)^2\dot{H}\dot{v}}{c_s^2}\right)=\dot{H}k^2 v \;,\quad \text{with } v \equiv \pi-\frac{\cal F}{k^2}\;.
\ee
We first solve equation \eqref{EoM v} subject to the boundary condition 
\be\label{v BC}
v(\rUV)=v_0\;, \quad v(r\to -\infty)=0\;,
\ee
This can be done using the same matching method introduced in Appendix \ref{eq:RigorousActionComputations}. Once the classical solution for $v$ (or $\pi_{\rm cl}$) is known, we can extract $A_{\rm cl}$ via \eqref{pidot eq}. 

Once again we would solve \eqref{EoM v} in the $z-$ coordinate, and here we focus on the low energy modes, with $k\ll z_2^{-1}$. Near the UV boundary $\zUV \le z \le z_1$, they are given, in series of $z$, by 
\begin{align}
\pi_{\rm cl}(\vec k, z)&=v_0+\frac{{\cal F}(\vec k)}{k^2}+\frac{v_0}{2\vep_0}\left(1-\frac{H_{\rm IR}}{H_{\rm UV}}\right)(k z)^2+\dots\;,\label{axial pi UV}\\
A_{\rm cl}(\vec k, z)&=\frac{2H_{\rm UV}{\cal F}(\vec k)}{k^2}+2H_{\rm UV}v_0\left(1-\frac{H_{\rm IR}}{H_{\rm UV}}\right)
\left(1+\frac{1}{4}(k z)^2\right)+\dots\;,\label{axial A UV}
\end{align}
where $\HUV=L_{\rm UV}^{-1}\;, \HIR=L_{\rm IR}^{-1}$.
On the other hand, in the deep IR region, $z\to \infty$, the classical solutions are approaching some asymptotic values: 
\begin{align}
\pi_{\rm cl}\sim \frac{{\cal F}(\vec k)}{k^2}+v_0(kz)^{1/2}e^{-kz}\;,\quad 
A_{\rm cl}\sim \frac{2H_{\rm IR}{\cal F}(\vec k)}{k^2}\;.
\end{align}
In what follows, we will focus on the case with $\CF=0$, so that both $A_{\rm cl}$ and $\pi_{\rm cl}$ vanish in the deep IR region.
The boundary value for $\pi_{\rm cl}$ and $A_{\rm cl}$ are not linearly independent in this case: 
\be\label{axial dependence}
 \pi_{\rm cl}(\vec k, \rUV)=v_0(\vec k)\;,\quad  A_{\rm cl}(\vec k, \rUV)=-2\left(H_{\rm IR}-H_{\rm UV}\right) \pi_{\rm cl}(\vec k, \rUV)\;.
\ee
Furthermore, for simplicity we can set the UV boundary value $\psi_{\rm cl}(\vec k, \rUV)=0$; this can always be done via a boundary diff.
 One advantage of working in this choice is that no IR regulator brane is needed, since all fields go to zero as $r\to -\infty$ (or $z\to +\infty$).

Thus the bulk action \eqref{Bulk Action} in the axial gauge becomes 
\begin{align}\label{bulk action axial}
S_{\rm axial}\simeq\mplthr&\int\!\rmd r \rmd^2 x\,
\Bigg[-\frac{a(r)^2}{4}(\dot{A}^2+\dot{A}\dot{\psi})
+\frac{a(r)^2\dot{H}\pi}{2}(2\dot{A}+\dot{\psi})
-\frac{a(r)^2\dot{H}\dot{\pi}^2}{2c_s^2}\nonumber\\
&\quad \quad
-\frac{\dot{H}}{2}(\pd \pi)^2
-a(r)^2\dot{H}^2\pi^2
-\frac{\rmd }{\rmd r}
\Bigg(a(r)^2\dot{H}\pi\left(1+A+\frac{\psi}{2}\right)
\Bigg)
\Bigg]\;.
\end{align}
Plugging in the classical solutions for $\pi_{\rm cl}$, $A_{\rm cl}$ and $\psi_{\rm cl}$ and we get an expression for the on-shell bulk action: 
\begin{align}\label{axial action onshell}
S_{\rm axial}^{\rm on-shell}[v_0]=\mplthr \int_{r=\rUV} \frac{\rmd^2 k}{(2\pi)^2}
\Bigg[&\frac{k^2}{2}\left(\frac{1}{H_{\rm UV}}-\frac{1}{H_{\rm IR}}\right)
\Bigg(
(H_{\rm UV}\pi_{\rm cl})^2- A_{\rm cl}H_{\rm UV}\pi_{\rm cl}
\Bigg)
-\frac{k^2 A_{\rm cl}^2}{8H_{\rm IR}}\nonumber\\
&\quad -a(r)^2\dot{H}\pi_{\rm cl}\left(1+A_{\rm cl}\right)
\Bigg]\;,
\end{align}
where it is understood that $A_{\rm cl}(\vec k,\rUV)$ and $\pi_{\rm cl}(\vec k,\rUV)$ are functions of $v_0$, given by \eqref{axial dependence}. We leave the total derivative term in \eqref{bulk action axial} in its original form, since it facilitates  discussion of the anomaly.

At first glance, the axial on-shell action in equation \eqref{axial action onshell} looks almost like what we want to match the boundary generating function $W_{\rm QFT}[\tau, \zeta_{(0)}]$, except for two subtleties: ({\it i}) there is an extra term, the second line of equation \eqref{axial action onshell}, in the bulk action, and ({\it ii}) the boundary value for $A_{\rm cl}$ and $\pi_{\rm cl}$ are not independent, so $S_{\rm axial}^{\rm on-shell}$ depends on one free parameter rather than two, as $W_{\rm QFT}[\tau, \zeta_{(0)}]$ does.

To understand those subtleties, we need recall the residual gauge freedom in the axial gauge.  It is well known that we haven't yet depleted the gauge freedom by demanding $N=1\;, N_i=0$ --- there is a residual gauge symmetry in the axial gauge preserving this choice, namely the coordinates transform as 
\be\label{residual transf axial gauge}
x^i\to x'^i=x^i-\pd_i\tau(\vec x) \int_{r_{\rm UV}}^r \frac{\rmd r_1}{a(r_1)^2}\;,\quad r\to r'=r+\tau(\vec x)\;,
\ee
while the field transformations are induced by
\be\label{field gauge transf}
\phi(\vec x, r)\to \phi'(\vec x',  r')=\phi(\vec x, r)\;,\quad 
h_{ij}(\vec x, r)\to h'_{ij}(\vec x', r')=\frac{\pd x^m}{\pd x'^i}\frac{\pd x^\ell}{\pd x'^j}h_{m \ell}+\frac{\pd \tau}{\pd x'^i}\frac{\pd \tau}{\pd x'^j}\;.
\ee
In the holographic computation, it is generally hard to deal with computationally a gauge transformation as a spacetime transformation, due to the presence of the regulator UV boundary: after a coordinate like \eqref{residual transf axial gauge}, an $r=$ constant surface becomes ${\vec x}-$dependent in the new coordinate. 

So instead we consider {\it internal} transformations induced by the spacetime gauge transformations
--- we only change fields according to \eqref{field gauge transf} at the {\it same} spacetime point while leave coordinates untouched. These transformations are 
\begin{align}\label{internal transf axial gauge}
\pi(x)\to \tilde{\pi}(x)&=\pi(x)-\sigma(\vec x)(1+\dot{\pi})+\pd_i \sigma(\vec x)\pd_i\pi(x)\int_{r_{\rm UV}}^r \frac{\rmd r_1}{a(r_1)^2}+\dots\;,\nonumber\\
A(x)\to \tilde{A}(x)&=A(x)-2H\sigma(\vec x)+\dots\;,\nonumber\\
\psi(x)\to \tilde{\psi}(x)&=\psi(x)+2\pd^2\sigma(\vec x)\int_{r_{\rm UV}}^r \frac{\rmd r_1}{a(r_1)^2}+\dots\;,
\end{align}
Here we work up to linear order in the gauge parameter $\sigma(\vec x)$, but, in principle, to all orders in fields, with $\dots$ denoting terms in higher order in the fields.  
We keep terms linear in  $\pi$ transformation for later discussions.  It is straightforward to check that the equations of motion in the axial gauge, \eqref{pidot eq}, \eqref{Adot eq} and \eqref{EoM psi}, are indeed unaltered under transformations \eqref{internal transf axial gauge}. The classical solutions for $\psi$ with different values of $\CF$ are related by such internal transformations.

We can use these internal transformations to change the boundary value for $\pi_{\rm cl}$ and $A_{\rm cl}$ to accommodate any boundary conditions imposed on the UV brane. For instance, if the boundary conditions are prescribed to be 
\be\label{UV bc axial}
A_{\rm cl}(\vec x, \rUV)=A_{(0)}(\vec x)\;,\quad \pi_{\rm cl}(\vec x, \rUV)=\pi_{(0)}(\vec x)\;,
\ee
we just need to specify $v_0=\frac{H_{\rm UV}}{H_{\rm IR}}\pi_{(0)}-\frac{ A_{(0)}}{2H_{\rm IR}}$, and then perform an internal gauge transformation with $\sigma(\vec x)=-\left(1-\frac{H_{\rm UV}}{H_{\rm IR}}\right)\pi_{(0)}(\vec x)-\frac{A_{(0)}(\vec x)}{2H_{\rm IR}}$. However, the transformations \eqref{internal transf axial gauge} will alter the IR behaviors of the classical solutions,  which will bring in unnecessary complication.  Instead we consider another set of slightly different internal gauge transformations, by promoting $\sigma(\vec x)$ to be an $r-$dependent function over the bulk.  We demand $\sigma(\vec x,r) $ has the properties that its $r-$dependence is weak near the UV boundary and it drops to zero sufficiently fast as $r\to -\infty$.  Therefore, these transformations will only preserve the conditions $N=1\;, N_i=0$ in the UV and IR region, but in the intermediate regime we are no longer in the axial gauge.

The bulk action, which is diff-invariant by construction, is {\it not} necessarily invariant under the internal gauge transformations alone; in general the variation of the bulk action yields a boundary term, if the gauge parameter does not vanish on the boundary.  In the case under consideration, we will see soon that this non-invariant piece is precisely the anomaly term for the boundary QFT. 

To be more explicit, we consider the variation of the our action $S_{\rm bulk}[g,\pi]$ under a generic transformation $g_{\mu\nu}\to g_{\mu\nu}+\Delta g_{\mu\nu}\,, \pi\to \pi+\Delta\pi$.  When the action is evaluated on on-shell field configurations, the variation is given by \cite{Balasubramanian:1999re, Brown:1992br}
\begin{align}
\Delta S_{\rm bulk}^{\rm on-shell}[g,\pi]&=-\frac{\mplthr}{2}\int_{r=r_{\rm UV}}\rmd^2 x\sqrt{h}\left(K^{ij}-K h^{ij}-L_{\rm UV}^{-1}h^{ij}\right)\Delta h_{ij}\nonumber\\
\quad &+\int_{r=r_{\rm UV}}\rmd^2 x \frac{\pd L_M}{\pd \dot{\pi}}\Delta \pi +\text{IR boundary terms}\;,
\end{align}
where $h_{ij}$ is the induced metric on the boundary. As promised, this variation is only sensitive to the change of fields on the boundary. The variation of the matter Lagrangian can be easily computed, with the aid of equation \eqref{Bulk Matter} and \eqref{eq:QDef}:
\begin{align}
\frac{\pd L_M}{\pd \dot\pi}\Big\vert_{r=\rUV}&=\sqrt{h}\sum_{n=1}^{\infty}\frac{1}{(n-1)!}M_n(r+\pi)Q^{n-1}\frac{2}{N}(1+\dot{\pi}-N^i\pd_i\pi)\Big\vert_{r=\rUV}\nonumber\\
&=2\sqrt{h}M_1(r+\pi)(1+\dot{\pi}),
\end{align}
where in the second equality, we have used the fact that the condition $N=1, N_i=0$ is maintained near the UV boundary, and that the coefficients $M_2=M_3=\dots =0$ there. Specifying 
\begin{align}
\Delta h_{ij}(\vec x, \rUV)&=a(\rUV)^2 \delta_{ij} \big(-2\HUV\sigma(\vec x, \rUV)\big)\;,\nn\\ 
\Delta \pi (\vec x, \rUV)&=-\big(1+\dot{\pi}(\vec x, \rUV)\big)\sigma(\vec x, \rUV)\;,
\end{align}
we find that the on-shell bulk action \eqref{axial action onshell}, under the internal transformations of the form \eqref{internal transf axial gauge} (with $\sigma$ acquiring an $r-$dependence), is shifted by 
\be\label{delta S v0}
\Delta_{\sigma}S_{\rm bulk}^{\rm on-shell}[v_0]=\mplthr (\HUV-\HIR)\int \frac{\rmd^2 k}{(2\pi)^2}k^2\sigma(\vec k,\rUV) v_0(-\vec k)\;.
\ee

Now we are in the position to compute the on-shell bulk action with independent boundary data $A_{(0)}$ and $\pi_{(0)}$.
We only work on the case in which the boundary values are infinitesmally away from equation \eqref{axial dependence}: 
\be
 A_{(0)}=-2\left(H_{\rm IR}-H_{\rm UV}\right) \pi_{(0)}+\delta f\;,\quad |\delta f|\ll 1
\ee
As we discussed earlier in this section, by choosing 
\begin{align}
v_0&=\frac{H_{\rm UV}}{H_{\rm IR}}\pi_{(0)}-\frac{ A_{(0)}}{2H_{\rm IR}}\;, \nn\\
\sigma(\vec x, \rUV)&=-\left(1-\frac{H_{\rm UV}}{H_{\rm IR}}\right)\pi_{(0)}(\vec x)-\frac{A_{(0)}(\vec x)}{2H_{\rm IR}}=-\frac{\delta f}{2H_{\rm IR}}\ll 1\;,
\end{align}
and combining equation \eqref{axial action onshell}and \eqref{delta S v0}
we can conclude that the on-shell action with the boundary data $A_{(0)}$ and $\pi_{(0)}$ must take the following form: 
\begin{align}\label{approx axial action onshell}
S_{\rm axial}^{\rm on-shell}[A_{(0)}, \pi_{(0)}]&=S_{\rm axial}^{\rm on-shell}[v_0]+\Delta_{\sigma}S_{\rm bulk}^{\rm on-shell}[v_0]\nn\\
&=\mplthr \int\!\frac{\rmd^2 k}{(2\pi)^2}
\Bigg[\frac{k^2}{2}\left(\frac{1}{H_{\rm UV}}-\frac{1}{H_{\rm IR}}\right)
\Bigg(
(H_{\rm UV}\pi_{(0)})^2- A_{(0)}H_{\rm UV}\pi_{(0)}
\Bigg)
-\frac{k^2 A_{(0)}^2}{8H_{\rm IR}}\Bigg]\nonumber\\
&\quad-\mplthr  \int_{r=\rUV}\rmd^2 x \,a(r)^2\dot{H}\pi_{\rm cl}\left(1+A_{\rm cl}\right)
\;,
\end{align}
with the understanding that $A_{(0)}$ and $\pi_{(0)}$ are completely free parameters in the above expression.

The first line of equation \eqref{approx axial action onshell} can be recast, up to quadratic order in perturbations,  into the form of 
\begin{align}\label{axial master equation}
S_{\rm axial}^{\rm on-shell}\supset\frac{c_{\rm UV}-c_{\rm IR}}{24\pi}\int\! \rmd^2 x
\Bigg((\pd \tau)^2+R^{(2)}\tau
\Bigg)
-\frac{c_{\rm IR}}{24\pi}\int\! \rmd^2 x\,(\pd \zeta_{(0)})^2\;,
\end{align}
where $\tau=-H_{\rm UV}\pi_{(0)}\;, \zeta_{(0)}=A_{(0)}/2$,  and $R^{(2)}=-2\pd^2 \zeta_{(0)}$ is the Ricci scalar constructed from the boundary metric $g_{(0)ij}=(1+2 \zeta_{(0)})\delta_{ij}$.  This is precisely equal to the low-energy generating function $W_{\rm QFT}[\tau, \zeta]$ given in Section \ref{sec:Flows}, with the dilaton field $\tau$ and the Weyl factor $\zeta_{(0)}$ being unrelated.

As was argued in Ref.~\cite{Komargodski:2011xv}, if the ``true'' Weyl transformation 
\be
\Delta \tau=\tilde{\sigma}\;,\quad \Delta \zeta=-\tilde{\sigma}\;,\quad \Delta \psi_{(0)}=0
\ee
were all that we consider, the boundary action, up to $\pd^2$ level,  should be given by equation \eqref{axial master equation}, without extra pieces that are exactly invariant. In our case, however, the (modified) Weyl transformation under consideration is just a remnant of the bulk (internal) gauge transformation \eqref{internal transf axial gauge}, which nonlinearly depends on fields. That is the reason why the term in the second line  of \eqref{approx axial action onshell}, which is invariant under the (full) internal gauge transformation, is also present in addition to the Wess-Zumino type term. 

In fact we can show straightforwardly that  the second line  of \eqref{axial action onshell} is invariant under \eqref{internal transf axial gauge} (taking into account the field-dependent piece):
\begin{align}
 \Delta\Bigg[-\int_{r=\rUV} \frac{\rmd^2 k}{(2\pi)^2}a(r)^2\dot{H}\pi_{\rm cl}\left(1+A_{\rm cl}\right)\Bigg]
 &=\int_{r=\rUV}\frac{\rmd^2 k}{(2\pi)^2}a(r)^2\dot{H}\sigma\Big[(1+\dot{\pi}_{\rm cl})(1+A_{\rm cl})+2\HUV \pi_{\rm cl}\Big]\nonumber\\
 &=\int_{r=\rUV}\frac{\rmd^2 k}{(2\pi)^2}a(r)^2\dot{H}\sigma\Big[\dot{\pi}_{\rm cl}+2\HUV \pi_{\rm cl}\Big]=0\;,
\end{align}
where in the second line we have used the classical solutions \eqref{axial pi UV} \eqref{axial A UV}, and have taken the limit $\vep_0 \to 0$.

In summary, we have computed the bulk on-shell action in an approximate axial gauge, --- satisfying $N=1\,, N_i=0$ only in the UV and IR region,  --- and it agrees with the boundary generating function $W_{\rm QFT}[\tau, \zeta_{(0)}]$. The variation of it under the induced internal transformation \eqref{internal transf axial gauge} (with $\sigma$ being $r-$ dependent ) reads
\be
\Delta S_{\rm axial}^{\rm on-shell}=\frac{c_{\rm UV}}{24\pi}\int\! \rmd^2 x\, \tilde{\sigma}\, R^{(2)}\;,
\ee
which is precisely the trace anomaly of the boundary QFT.

\bibliographystyle{utphys}
\bibliography{EFTHolographicRGBib}

 \end{document}